\documentstyle{article}

\begin{document}
\large

\smallskip
\centerline {  Multiple Hamiltonian Structures  }
\centerline {   for   Toda-type systems }
\vskip 1  cm
\centerline { Pantelis A. Damianou }
\centerline { Department of Mathematics and Statistics } 
\centerline  { University of Cyprus}
\centerline { P. O. Box 537, Nicosia, Cyprus }
\vskip 2 cm
\centerline { \bf {ABSTRACT }}
\bigskip
{\it Results on the finite nonperiodic  Toda lattice are extended 
 to some generalizations of the system: The relativistic Toda lattice,
  the generalized Toda lattice associated with  
simple Lie groups and the full Kostant-Toda lattice.  
   The areas  investigated, include  master symmetries, recursion operators, higher    Poisson
brackets, invariants, and group symmetries for the systems.
A survey of previous work on the classical Toda lattice is also included.}


\vfill
\eject    

\renewcommand{\thesection}{\Roman{section}}
\section   {    INTRODUCTION}
\vskip 2 cm

This paper can be divided into  two parts. First, we survey some previous work on the Toda
lattice dealing with the Poisson Geometry and  algebraic structure of the system (section III). The second part contains
 new results on some systems generalizing the finite nonperiodic Toda lattice (sections IV, V, VI). This system has been studied 
extensively and therefore we   will examine only some aspects of the Toda lattice which   are shared by most 
 integrable systems.
The list of topics includes  bi-Hamiltonian structure, recursion operators,  symmetries, master symmetries,
 Lax formulations, Poisson and symplectic Geometry. This area of integrable systems  has been studied extensively for 
infinite dimensional systems such as the KdV,  Burgers,  Kadomtsev-Petviashvili, Benjamin-Ono
 equations  and many more. The Toda lattice is the only finite dimensional system where these ideas have been worked out
in detail.  In this paper we will extend results about the finite nonperiodic Toda lattice to some 
generalizations of the system: The relativistic Toda lattice, the generalized Toda lattice associated with 
simple Lie groups and the full Kostant-Toda lattice.

\smallskip
The natural setting for Hamiltonian systems is on symplectic manifolds. These  are  Poisson  manifolds 
 whose Poisson structure is locally isomorphic to  the standard one on ${\bf R}^{2N}$.  A Poisson 
structure on a manifold $M$ may be defined as contravariant tensor field (bivector) $\pi$ for which the
Poisson bracket on $C^{\infty}(M)$,  $\{f, g \} = \langle \pi, df \wedge dg \rangle $  satisfies
 the Jacobi identity.  When $\pi$ has full rank, the dimension of $M$ is even and the Poisson structure is 
symplectic. Darboux's theorem provides coordinates which make the structure locally isomorphic to the 
standard symplectic bracket  on ${\bf R}^{2N}$. To define  a Hamiltonian system  we consider  ${\bf R}^{2N}$  with coordinates 
 $(q_1, \dots , q_N, p_1, \dots, p_N)$, and  the
standard symplectic bracket 
\begin{equation}
\{ f, g \} = \sum_{i=1}^N ( {\partial f \over \partial q_i} {\partial g 
\over \partial p_i} - {\partial f \over \partial p_i} { \partial g \over
\partial q_i} )   \ .
\end{equation}
Let $H: {\bf R}^{2N} \to {\bf R}$ be a smooth function. Hamilton's equations 
are the differential equations 

\begin{equation}
\dot q_i={\partial H \over \partial p_i}, \ \ \ \ \ \ \  \dot p_i =-{\partial H \over \partial q_i}  \ .
\end{equation}

Using the symplectic bracket (1), Hamilton's equations take the simple form
\begin{equation}
\dot F =\{ F,H \} \ .
\end{equation}
The condition $\dot F=0$ is equivalent to the condition $\{ F, H \}=0$. Such a function is
called a constant of motion (or first integral).

\smallskip
Smooth functions $f_1, \dots, f_r$ on a manifold $M$ are called independent,
 if $df_1 \wedge \dots  \wedge df_r \not= 0$, except possibly on a submanifold of 
smaller dimension.  The functions $f_1, \dots, f_r$ are said to be 
mutually in involution if $\{f_i, f_j \}=0,  \ \ \  \forall \, i, j$. A
Hamiltonian system is called  integrable if there exists a family $f_1, \dots,
f_r$ of independent $C^{\infty}$ functions mutually in involution. The 
Hamiltonian is one of these functions (or a function of these functions).  
The systems we consider will be integrable.

\smallskip
 The equations  for the Toda systems in consideration  will be  written in the form

\begin{equation}
\dot L(t)=[B(t), \,  L(t)]  \ . 
\end{equation}
The pair of matrices $L$,  $B$ is  known as a Lax pair. In the case of the finite
nonperiodic  Toda lattice $L$ is a symmetric tridiagonal matrix and $B$ is 
  the projection onto the skew-symmetric part  in the 
decomposition of $L$ into skew-symmetric plus lower triangular.  In  the case of the other 
generalized Toda systems the matrix $L$ will lie in some  Lie algebra  and 
$B$ will again be obtained from $L$ by some projection associated with  a decomposition
of the Lie algebra. The decomposition plays an important role in the solution of the
equations by factorization. 

\smallskip
\noindent
A Lax equation implies that the  traces of powers of $L$ are constants of
motion.  $\dot L = [ B, L]$ implies   ${ (L^2 \dot )} = [B, L^2]$,
 and similarly  ${ (L^k \dot )} = [B, L^k]$.  Therefore,

\begin{equation}
( {\rm trace} \ L^k \dot ) = { \rm trace \ }  {  (L^k \dot )} = { \rm trace \ } [B, L^k]=0 \ .  
\end{equation}
 This calculation shows that the eigenvalues of $L$ do not evolve with time.  Therefore, such
 systems will be integrable.

\smallskip
In the case of Toda lattice the Lax equation is obtained by the use of a transformation due
 to H. Flaschka \cite{flaschka1}  which changes the original $(p,q)$ variables to new reduced variables $(a,b)$.
  The  symplectic bracket in the variables $(p,q)$ transforms to a degenerate Poisson bracket in 
the variables $(a,b)$. This linear bracket   is an example of a Lie-Poisson bracket. 
The functions $H_n={ 1 \over n} \, {\rm tr}\, L^n$ are in involution.  A 
Lie algebraic interpretation of this bracket can be found in \cite{kostant}.  We  denote this bracket by
 $\pi_1$. A quadratic Toda bracket, which we call $\pi_2$ appeared in a paper of Adler  \cite{adler}.
 It is a Poisson bracket in which the Hamiltonian vector field generated by $H_1$ is the 
same as the Hamiltonian vector field generated by $H_2$ with respect to the $\pi_1$ bracket.
  This is  an example of a bi-Hamiltonian system,  an idea introduced by Magri \cite{magri}.  A cubic bracket
 was found by Kupershmidt \cite{kupershmidt} via the infinite Toda lattice. We found the explicit 
formulas for both the quadratic and cubic brackets in some lecture notes by H. Flaschka.
The Lenard  relations (73)   are also in these notes.

 In  \cite{damianou1} we used master symmetries to generate nonlinear
Poisson brackets for the Toda lattice.  In essence, we have  an example of a system  
 which is not only bi-Hamiltonian but it can actually be given $N$ different
 Hamiltonian formulations with $N$ as large as we please. The first three Poisson brackets
are precisely the linear, quadratic and cubic brackets  we mentioned above, but one can 
use the master symmetries to produce an infinite hierarchy of brackets.
If a system is bi-Hamiltonian and one of the brackets is symplectic,  one can find a recursion 
operator by inverting the  symplectic  tensor.  The recursion operator is then applied to
the initial symplectic bracket  to produce an infinite sequence.  However, in the case of
Toda lattice (in  Flaschka variables $(a,b)$) both operators are non-invertible and therefore this method
fails.   The absence of a recursion operator for the finite Toda lattice is also mentioned in 
Morosi and Tondo \cite{morosi} where a Ninjenhuis tensor for the infinite Toda lattice is calculated.
 Recursion operators were introduced  by Olver \cite{olver1}. 
Master  symmetries were first introduced by Fokas and Fuchssteiner in \cite{fokas1}  in 
connection with the Benjamin-Ono Equation. Then  W. Oevel and B. Fuchssteiner \cite{oevel1}
  found master symmetries for the  Kadomtsev-Petviashvili equation. 
 Master symmetries for  equations in $1+1$,  like the KdV, are discussed in  Chen, 
  Lee and Lin \cite{chen}  and in  Fokas \cite{fokas2}. The   general theory of master symmetries is discussed
  in Fuchssteiner \cite{fuchssteiner}.   In the case of Toda equations
 the master symmetries map invariant functions  to other invariant functions.
 Hamiltonian vector fields are also preserved. New Poisson brackets are generated 
by using Lie derivatives in the direction of these vector fields and they satisfy 
interesting deformation relations. 
Another approach, which explains these relations is adopted in  Das and Okubo \cite{das},  and Fernandes 
 \cite{fernandes}. 
 In principle, their  method is  
general and may work for other finite dimensional systems as well. The procedure is the following:
One defines a second Poisson bracket in the space of canonical variables 
$(q_1, \dots, q_N, p_1,\dots, p_N)$.
This gives rise to 
a recursion operator. The presence of a conformal symmetry
 as defined in Oevel \cite{oevel2} allows one, by using the recursion 
 operator, to generate an infinite sequence of master symmetries. These, in 
turn, project to the  space of the new variables $(a,b)$ to
produce a sequence of master symmetries in the reduced space.  This approach is
discussed at the end of section III.

\smallskip
The Toda lattice has been generalized in several directions:  Kostant  \cite{kostant} and
Bogoyavlensky \cite{bogoyavlensky} 
 generalized the system to the tridiagonal coadjoint orbit of the Borel subgroup of an 
arbitrary simple Lie group. Therefore,  for each simple Lie group there  is a 
corresponding  mechanical system of Toda type. 

 Another generalization  is due to    Deift, Li, Nanda and
 Tomei \cite{deift} who showed that the system remains integrable when $L$ is replaced 
 by a full (generic) symmetric $n \times n$ matrix.

 Another variation is the 
full Kostant-Toda lattice which was studied by S. Singer, N. Ercolani and H. Flaschka  
 \cite{singer}, \cite{ercolani}, \cite{flaschka3}.  In this case,  the matrix
$L$ in the Lax equation is the sum of a lower triangular plus  a regular nilpotent matrix.

\smallskip
 Finally, there is a  relativistic Toda lattice  which was introduced by Ruijsenaars \cite{ruijsenaars}.  
 The non-relativistic Toda lattice can be thought as a limiting case of this system.

\smallskip
 In section II we present the necessary background on Poisson manifolds, bi-Hamiltonian systems and
 master symmetries. In the spirit of this special issue we define Cohomology of Lie algebras and a
 Cohomology on  Poisson manifolds due to Lichnerowicz \cite{lichnerowicz}.  Most of the results   in this 
paper can be considered  as statements
about the Cohomology of Poisson Lie algebras.

\smallskip
Section III is a review of  the classical finite nonperiodic Toda lattice. This system was investigated in 
  \cite{flaschka1},  \cite{toda}, \cite{henon},  \cite{flaschka4}, \cite{moser}.  We define the quadratic and cubic
Toda brackets and show that they satisfy certain  Lenard-type relations. We briefly describe the construction of 
master symmetries and the new Poisson brackets as in \cite{damianou1}.  Propositions 2-5 have been
known for some time \cite{damianou2},  but had not been published before. Theorem 7  which shows the connection  with 
group symmetries appeared in \cite{damianou3}. 

\smallskip

In section IV, some results on the non-relativistic Toda lattice are extended to the 
case of the relativistic Toda Systems.  The main new result is the  hierarchy of
higher Poisson brackets which also exists in the finite nonperiodic 
Toda lattice.    We were recently informed by Walter Oevel
 about reference \cite{oevel3},
 where  the first three local Hamiltonian structures are found using different methods.

\smallskip
In section V we define some integrable systems associated with  simple Lie groups. They have been 
considered by Kostant \cite{kostant},  Bogoyavlensky \cite{bogoyavlensky} and Olshanetsky and Perelomov \cite{olshanetsky}.
 We present in detail the systems of
type $B_n$. We show that they are bi-Hamiltonian and we also construct a recursion operator. These
results are new. Similar results hold for other semisimple Lie algebras. We  checked small 
dimensions,  but  we do not have complete general results yet.

\smallskip
In section VI    we define  Poisson structures   which lead to
 a tri-Hamiltonian formulation for the full Kostant-Toda
lattice.  In addition,  master symmetries  
 are constructed and they are used to generate the nonlinear Poisson brackets
and other invariants.
Various deformation relations are investigated. The results have been announced in \cite{damianou4}.
 
\bigskip
\section  { BACKGROUND}

\subsection*{ The Schouten bracket}

 We  list some properties of the Schouten bracket following Lichnerowicz \cite{lichnerowicz}.
Let $M$ be a $C^{\infty}$  manifold, $N=C^{\infty}(M)$ the algebra of $C^{\infty}$ real valued 
functions on $M$. 
 A contravariant, antisymmetric tensor of order $p$ will be called a $p$-tensor for short. These
 tensors form a superspace endowed with a  Lie-superalgebra
 structure via the {\it Schouten bracket}.  

\smallskip
The Schouten bracket assigns to each $p$-tensor $A$, and $q$-tensor $B$, a
 $(p+q-1)$-tensor, denoted by $[A,B]$.
For $p=1$ we have $[A,B]=L_A B$ where $L_A$ is the Lie-derivative in  the
direction of the vector field $A$.
\bigskip
 The bracket satisfies:

\smallskip
\noindent
{\it i)} \begin{equation}
  [A,B]= (-1)^{pq} [B,A]
\end{equation}
\smallskip
\noindent
{\it ii)} If $C$ is a $r$-tensor
   \begin{equation}
(-1)^{pq} [[B,C], A]+ (-1)^{qr} [[C,A],B] +(-1)^{rp} [[A,B],C]=0
\end{equation}
\smallskip
\noindent
{\it iii)} \begin{equation}
[A, B\wedge C]= [A,B]\wedge C + (-1)^{pq+q} B \wedge [A,C]  \ .
\end{equation}

\bigskip

\subsection*{Poisson Manifolds}  

A { \it Poisson structure} on $M$ is a bilinear form, called the { \it Poisson bracket}  $\{\ , \ \} : \   
   N \times N \to N$ such that

\smallskip
\noindent
{\it i)} \begin{equation}
     \{ f, g \} = - \{ g, f \}
\end{equation}
\smallskip
\noindent
{\it ii)} \begin{equation}
\{ f, \{ g, h\} \} + \{ g, \{h, f \}\} + \{ h, \{ f, g \} \}=0
\end{equation}
\smallskip
\noindent
{\it iii)}\begin{equation}
 \{ f, gh \} = \{ f, g \} h + \{f, h \} g 
\end{equation}
Properties {\it i)} and {\it ii)} define a Lie algebra structure on $N$. {\it ii)} is called 
the Jacobi identity and {\it iii)} is the analogue of Leibniz rule from calculus. A  {\it Poisson 
manifold} is a manifold $M$ together with a Poisson bracket $\{ \ , \ \}$.

\smallskip
 To a  Poisson bracket  one can associate a 2-tensor $\pi$ such that
 \begin{equation}
\{ f, g \}= \langle \pi , df \wedge dg \rangle \ .
\end{equation}
Jacobi's identity is equivalent to the condition $[ \pi,\pi]=0$ where $[ \ , \ ]$ is the 
Schouten bracket. Therefore, one could define a Poisson manifold by specifying  a pair 
$(M,\pi)$  where $M$ is a manifold and $\pi$ a 2-tensor satisfying $[\pi, \pi]=0$. In 
local coordinates $(x_1, x_2, \dots, x_n)$,   $\pi$ is given by 
\begin{equation}
\pi=\sum_{i,j} \pi_{ij} { \partial \over \partial x_i} \wedge { \partial \over \partial x_j} 
\end{equation}
  
and 
\begin{equation}
\{ f, g \}= \langle \pi , df \wedge dg \rangle =\sum_{i,j} \pi_{ij} { \partial f \over \partial x_i}
 \wedge { \partial  g \over \partial x_j} \ .
\end{equation}
In particular $ \{ x_i, x_j \} =\pi_{ij}(x)$.  Knowledge of the Poisson matrix $(\pi_{ij})$ is sufficient to 
 define the bracket of arbitrary functions. The rank of the matrix $( \pi_{ij})$
at a point $x \in M$ is called the {\it rank } of the Poisson structure at $x$. 

\smallskip
A function $F : M_1 \to M_2 $ between two Poisson manifolds is called
 a {\it Poisson mapping} if
\begin{equation}
\{ f \circ F, g \circ F \}_1 = \{ f, g \}_2 \circ F 
\end{equation}
for all $f, g \in  C^{ \infty}(M_2)$. In terms of tensors,  $F_* \pi_1=\pi_2$.
 Two Poisson manifolds are called { \it  isomorphic}, if there exists a diffeomorphism
 between them which is a Poisson mapping. 

\smallskip
The Poisson bracket  allows one to associate a vector field to each element
 $f \in N$.  The vector field $\chi_f$ is defined by the formula
\begin{equation}
\chi_f (g) = \{f, g\} \ .
\end{equation}
It is called the {\it Hamiltonian vector field generated by } $f$. In terms of the
Schouten bracket 
\begin{equation}
\chi_f = [ \pi, f] \ .
\end{equation}
Hamiltonian vector fields are {\it infinitesimal automorphisms}  of the Poisson 
 structure. These are vector fields $X$ satisfying $L_X \pi=0$. In the
case of Hamiltonian vector fields we have
\begin{equation}
L_{\chi_f} \pi =[\pi, \chi_f]=[\pi, [\pi, f]]=- 2 [[\pi, \pi], f]=0 \ .
\end{equation}

\smallskip
The Hamiltonian vector fields form a Lie algebra and in fact
\begin{equation}
[\chi_f, \chi_g] = \chi_  { \{f,g\}} \ .
\end{equation}
So, the map $f \to \chi_f$ is a Lie algebra homomorphism. 

\smallskip
The Poisson structure defines a bundle map $\pi^* : T^*M \to TM$ such
that
\begin{equation}
\pi^* (df) = \chi_f \ .
\end{equation}
The rank of the Poisson structure at a point $x \in M$ is the rank of 
$\pi_x^* : T_x^*  M \to T_x M$. 

\smallskip
The functions in the center of $N$ are called  {\it Casimirs}. It is the set 
of functions $f$ so that $\{f, g \}=0$  for all $g \in N$. These are functions
which are constant along the orbits of Hamiltonian vector fields. The 
differentials of these functions are in the kernel of $\pi^*$. In terms of
 the Schouten bracket a Casimir satisfies $[\pi, f]=0$. 

\smallskip
Given a function $f$, there is a reasonable algorithm for constructing 
a Poisson bracket in which $f$ is a Casimir. One finds two vector fields
 $X_1$ and $X_2$ such that $L_{X_1}f=L_{X_2}f =0$.  If in addition $X_1$,
 $X_2$ and $[X_1, X_2]$ are linearly dependent, then $X_1 \wedge X_2$ is
a Poisson tensor and $f$ is a Casimir in this bracket. In fact
\begin{equation}
[f, X_1 \wedge X_2]=[f,X_1] \wedge X_2 - X_1 \wedge [f,X_2]=0 \ .
\end{equation}

\smallskip
More generally, there is  a formula due to H. Flaschka and T. Ratiu which gives 
locally a Poisson bracket when the number of Casimirs is 2 less than the
 dimension of the space.
  Let $f_1,f_2,\dots, f_r$ be functions on ${\bf R}^{r+2} $.
Then the formula
\begin{equation}
 \omega \{ g, h \} =df_1 \wedge \dots \wedge df_r \wedge dg \wedge dh 
\end{equation}
where $\omega$ is a non-vanishing $r+2$ form, defines a Poisson bracket on 
${\bf R}^{r+2}$ and the functions $f_1, \dots, f_r $ are Casimirs.

\smallskip
Multiplication of a Poisson bracket by a Casimir gives another Poisson 
bracket. Suppose $[\pi, \pi]=0$ and $[\pi, f]=0$. Then
\begin{equation}
[f\pi, f\pi]=f \wedge [f, \pi] \wedge \pi+ f \wedge \pi \wedge [\pi,f]+
f^2 [\pi,\pi]=0 \  .
\end{equation}

\bigskip
\subsection*{Examples}

 The most basic examples of Poisson brackets are the symplectic and Lie-Poisson
brackets.

\bigskip
\noindent

{\it i)}\ { \bf Symplectic manifolds:} A {\it symplectic manifold} is a pair $(M^{2n}, \omega
)$ where $M^{2n}$ is an even dimensional manifold and $\omega$ is a closed, 
non-degenerate  two-form.
The associated isomorphism 
\begin{equation}
 \mu : TM \to T^* M 
\end{equation}
extends naturally to  a tensor bundle isomorphism still denoted by $\mu$. Let $\lambda= \mu^{-1}$,
 $f \in N$ and let $\chi_f= \lambda(df)$ be the corresponding Hamiltonian vector field. 
  The symplectic bracket is given by

\begin{equation}
   \{ f, g\}= \omega ( \chi_f , \chi_g )  \ .
\end{equation}
\smallskip
\noindent
In the case of ${ \bf R}^{2n}$,  there are  coordinates $(x_1,\dots, x_n, y_1,\dots, y_n)$, so that

\begin{equation}
\omega = \sum_{i=1}^n dx_i \wedge dy_i
\end{equation}
\smallskip
\noindent
and the Poisson bracket is the standard one (1).

\bigskip
\noindent
{\it ii)}\ {\bf Lie Poisson :}  Let $M=\cal{ G}^*$ where ${\cal G}$ is a Lie algebra. For
$a \in {\cal G}$ define the function $\Phi_a $ on ${\cal G}^*$ by 
\begin{equation}
 \Phi_a (\mu ) = \langle a , \mu \rangle 
\end{equation}
\smallskip
\noindent
where $\mu \in {\cal G}^*$ and $\langle \ , \  \rangle $  is the pairing
between ${\cal G}$ and ${\cal G}^*$. Define a bracket on ${\cal G}^*$ by
\begin{equation}
 \{ \Phi_a, \Phi_b \} = \Phi_{[a,b]} \ .
\end{equation}
\smallskip
\noindent
This  bracket is easily extended to arbitrary $C^{\infty }$ functions
on ${\cal G}^*$.  The bracket of linear functions is linear and every linear
bracket is of this form,  i.e.,   it is associated with a Lie algebra.

\bigskip
\subsection*{Local theory}

In his paper \cite{weinstein}  A. Weinstein proves the so-called
 ``splitting theorem'',  which describes the local behavior of Poisson
 manifolds.

\smallskip
\newtheorem{theorem}{Theorem}
 \newtheorem{proposition}{Proposition}
\begin{theorem}
 Let $x_0$ be a point in a Poisson manifold $M$. Then
near $x_0$, $M$ is isomorphic to a product $S \times N$ where $S$ is symplectic,
  $N$ is a Poisson manifold, and the rank of $N$ at $x_0$ is zero.
\end{theorem}
\noindent
$S$ is called the {\it symplectic leaf through } $x_0$ and $N$ is called the 
 {\it transverse Poisson structure at } $x_0$. $N$ is unique up to isomorphism.

\smallskip
So, through each point $x_0$ passes a symplectic leaf $S_{x_0}$ whose dimension
equals the rank of the Poisson structure  on $M$ at $x_0$. The bracket
on the transverse manifold $N_{x_0}$ can be calculated using {\it Dirac's constraint
bracket formula}. For more details, see Oh \cite{oh}.

\bigskip
\noindent
\begin{theorem}
  Let $x_0$ be a point in a Poisson manifold $M$
and let $U$ be a neighborhood of $x_0$ which is isomorphic to  a product
$S\times N$ as in Weinstein's splitting theorem. Let $p_i$, $ i=1, \dots,
2n$ be functions on $U$ such that
\begin{equation}
 N = \{ x\in U \, \vert \,  p_i (x) = {\rm constant} \} \ .
\end{equation}
\noindent
Denote by $P = P_{ij} = \{ p_i, p_j \}$ and by $P^{ij}$
the inverse matrix of $P$. Then the bracket formula for the
transverse Poisson structure on $N$ is given as follows:
\begin{equation}
\{ F, G \}_N (x) = \{ \hat{F}, \hat{G} \}_M (x) + \sum_{i,j}^{2n}
\{ \hat{F}, p_i \}_M (x) P^{ij } (x) \{ \hat{G}, p_j \}_M (x)
\end{equation}

\noindent
for all $x \in N$, where $F$, $G$ are functions on $N$ and $\hat{F}$, $\hat{G}$
are extensions of $F$ and $G$ to a neighborhood of $M$. Dirac's formula depends
only on $F$, $G$, but not on the extensions $\hat{F}$, $\hat {G}$.
\end{theorem}

\bigskip
\subsection*{Cohomology}
Cohomology of Lie algebras was introduced by Chevalley and Eilenberg in \cite{chevalley}.
Let $\cal G$ be a Lie algebra and let $\rho$ be a representation of $\cal G$ with representation
space $V$. A $q$-linear skew-symmetric mapping of $\cal G$ into $V$ will be called a 
$q$-dimensional $V$-cochain. The $q$-cochains form a space $C^q(\cal G, V)$. By definition,
$C^0 (\cal G,V)=V$.

\smallskip
 We define a coboundary operator
$ \delta = \delta _q   :  C^q ( {\cal G}, V)  \to  C^ {q+1}  ( {\cal  G} , V)$  by the formula

\begin{equation}
\begin{array}{rcl}
(\delta f)(x_0, \dots, x_q) &= &\sum_{i=0}^q (-1)^q \rho(x_i) f(x_0, \dots, \hat x_i, \dots, x_q)+ \\
                             && \\
  & &\sum_{i<j} (-1)^{i+j} f([x_i, x_j], x_0, \dots, \hat x_i, \dots, \hat x_j, \dots, x_q)
\end{array}
\end{equation}
where $f \in C^q ({\cal G},V)$ and $x_0, \dots, x_q \in {\cal G}$.
 As can be easily checked $\delta_{q+1} \circ \delta_q=0$ so that
$\{ C^q ({\cal G}, V), \delta_ q \}$ is an algebraic complex. Define $Z^q({\cal G}, V)$ the space
of $q$-cocycles as the kernel of $\delta : C^q \to C^{q+1} $ and the space $B^q({\cal G}, V)$ of
$q$-coboundaries as the image $\delta C^{q-1}$. Since $\delta \delta =0$ we can define
\begin{equation}
H^q ({\cal G}, V) = {  Z^{q} ({ \cal G},V) \over B^{q} ({\cal G},V) } \ .
\end{equation}

\smallskip
   
 Lichnerowicz \cite{lichnerowicz} considers the following cohomology defined on the tensors of a Poisson manifold.
Let $(M, \pi)$ be a Poisson manifold. If we set $B=C=\pi$ in (7) we get
\begin{equation}
[\pi, [\pi, A]]=0 
\end{equation}
for every tensor $A$. Define a coboundary operator $\partial_{ \pi}$  which assigns to each $p$-tensor 
$A$,  a $(p+1)$-tensor $\partial_{\pi} A$ given by
\begin{equation}
\partial_{\pi}A=-[\pi, A] \ .
\end{equation}
We have $\partial_{\pi}^2 A=[\pi, [\pi,A]]=0$ and therefore $\partial_{\pi}$ defines a cohomology.
 An element $A$ is a $p$-cocycle if $[\pi, A]=0$. An element $B$ is a $p$-coboundary if $B=[\pi,C]$,
 for some $(p-1)$-tensor $C$. Let 
\begin{equation}
Z^n(M, \pi) = \{ A \in T_n \, \vert \,  [\pi, A]=0  \}
\end{equation}
and
\begin{equation}
B^n(M, \pi)= \{ B \,  \vert \,  B=[\pi, C] \  \ C \in T_{n-1} \} \ .
\end{equation}
The quotient 
\begin{equation}
H^n(M, \pi)= {  Z^n(M, \pi) \over B^n(M, \pi) }
\end{equation}
is the $n$th cohomology group.

\bigskip
\noindent

\begin{proposition}
\smallskip

Let $(M, \pi_1)$, $(M, \pi_2)$ two Poisson structures on $M$. The following are equivalent:

\smallskip
\noindent
{\it i)} $\pi_1 +\pi_2$ is Poisson.

\smallskip
\noindent
{\it ii)} $[\pi_1, \pi_2]=0$.

\smallskip
\noindent
{\it iii)} $\partial_{\pi_1} \, \partial_{\pi_2}= -  \partial_{\pi_2} \, \partial_{\pi_1}$.

\smallskip
\noindent
{\it iv)} $\pi_1 \in Z^2 (M, \pi_2)$,  $\pi_2 \in Z^2(M, \pi_1)$. 

\end{proposition}

\bigskip
Two tensors which satisfy the equivalent conditions are said to form a {\it Poisson pair}  on $M$. The
corresponding Poisson brackets are called {\it compatible}.

\bigskip
\newtheorem{lemma}{Lemma}
\begin{lemma}
Suppose $\pi_1$ is Poisson and $\pi_2 =  L_{X}  \pi_1 =-\partial_{\pi_1} X $ for some vector field $X$.     
 Then $\pi_1$ is compatible with $\pi_2$.
 \end{lemma}

\smallskip
{\bf Proof:} 
\begin{equation}
[\pi_1, \pi_2]=[\pi_1, -[\pi_1 , X]]= -\partial_{\pi_1} \partial_{\pi_1} X=0  \ \ \ \ \ \ \ \ \   \Box   \ .
\end{equation}

\smallskip
\noindent
 If  $\pi_1$  is 
symplectic, we call the Poisson pair non-degenerate.  If we assume a non-degenerate pair we make
the following definition: The {\it recursion operator} associated with a non-degenerate pair is the
$(1,1)$-tensor ${\cal R}$ defined by
\begin{equation}
{\cal R}=\pi_2  \pi_1^{ -1}. 
\end{equation}

\bigskip
A  {\it bi-Hamiltonian system} is defined by specifying two Hamiltonian functions $H_1$, $H_2$ satisfying:
\begin{equation}
X=\pi_1  \,  \nabla H_2 = \pi_2 \, \nabla  H_1 \ .
\end{equation}
We have the following result due to Magri \cite{magri}  

\smallskip
\noindent

\begin{theorem}
 Suppose we have a bi-Hamiltonian system on a manifold $M$, whose first 
cohomology group is trivial. Then there exists a hierarchy of mutually commuting 
functions $H_1, H_2, \dots $,  all in involution with respect to both brackets. They
 generate mutually commuting bi-Hamiltonian flows $\chi_i$, $i=1,2,\dots$, satisfying the
Lenard recursion relations 
\begin{equation}
\chi_{i+j}= \pi_i \,   \nabla H_j  \ ,
\end{equation}
where $\pi_{i+1}={\cal R}^i \pi_1$ are the higher order Poisson tensors.  

\end{theorem}

\bigskip
Let ${ \cal G}$ be a Lie algebra and consider the Lie-Poisson manifold 
  ${\cal G}^*$.  Define a 
representation $\rho$ of ${\cal G}$ with values in $C^{\infty} ({\cal G}^*)$ by
\begin{equation}
\rho (x_i) f= \sum_{j,k} c_{ij}^k { \partial f \over \partial x_j}  
\end{equation}
where $x_i$ denotes coordinates on ${\cal G}^*$ and at the same time elements of a basis for
${\cal G}$. In other words,  $\rho(x_i)f =\{ x_i, f \}$,  where the bracket is the Lie-Poisson
 bracket on ${\cal G}^*$. We denote the $n$th cohomology group of ${\cal G}$ with respect
to this representation by
\begin{equation}
H^n ({\cal G}, C^{\infty}( {\cal G}^* )) \ .
\end{equation}
We have the following result:

\smallskip
\noindent

\begin{theorem}
\begin{equation}
H^n ({\cal G}^*, \pi) \cong H^n ({\cal G}, C^{\infty}( {\cal G}^* )) \ .
\end{equation}
\end{theorem}
The proof can be found in \cite{koszul} or \cite{damianou2}.

\bigskip
\subsection*{Master Symmetries}
\bigskip
We recall the definition and basic properties of master symmetries following
 Fuchssteiner \cite{fuchssteiner}. 

\smallskip
Consider a differential equation on a manifold $M$, defined by a vector field $\chi$.
 We are mostly interested in the case where $\chi$ is a Hamiltonian vector field. A
 vector field $Z$ is a  {\it symmetry} of the equation  if 
\begin{equation}
[Z, \chi]=0 \ .
\end{equation}
If $Z$ is time dependent, then a more general condition is
\begin{equation}
{ \partial Z \over \partial t} + [ Z, \chi]=0  \ .
\end{equation}
 A vector field $Z$ will be called a master symmetry if 
\begin{equation}
[[Z, \chi], \chi]=0 \ ,
\end{equation}
but
\begin{equation}
[Z, \chi] \not= 0  \ .
\end{equation}
Suppose that we have a bi-Hamiltonian system defined by the  Poisson tensors $J_0$, $J_1$ and
the Hamiltonians $h_0$, $h_1$. 
 Assume that $J_0$ is symplectic.  We define
the recursion operator ${\cal R} = J_1 J_0^{-1}$,  the higher flows 
\begin{equation}
\chi_i = {\cal R}^{i-1} \chi_1 \ ,
\end{equation}
and the higher order Poisson tensors
\begin{equation}
J_i = {\cal R}^i J_0 \ .
\end{equation}    
Master symmetries preserve constants of motion, Hamiltonian vector fields and
generate hierarchies of Poisson structures.  For a non-degenerate 
bi-Hamiltonian system, master symmetries can be generated using a method due to 
W. Oevel \cite{oevel2}.

\begin{theorem}
 Suppose that   $Z_0$ is a conformal symmetry for both  $J_0$, $J_1$ and $h_0$. 
i.e.,  for some scalars $\lambda$, $\mu$, and $\nu$ we have
\begin{equation}
L_{Z_0} J_0= \lambda J_0,  \ \ \ \ \ L_{Z_0} J_1 =  \mu J_1, \ \ \ \ \ \ \ \ L_{Z_0} h_0 = \nu h_0 \ .
\end{equation}
 Then the vector fields 
\begin{equation}
Z_i = {\cal R}^i Z_0
\end{equation}
are master symmetries and  we have

\smallskip
\noindent
(a) 
\begin{equation}
[Z_i, \chi_j] = (\mu + \nu + (j-1) (\mu- \lambda)) \chi_{i+j} 
\end{equation} 

\smallskip
\noindent
(b)
\begin{equation}
[Z_i, Z_j]= (\mu - \lambda) (j-i) Z_{i+j} 
\end{equation}

\smallskip
\noindent
(c)

\begin{equation}
L_{Z_i} J_j = (\mu +(j-i-1) (\mu -\lambda)) J_{i+j} 
\end{equation}

\end{theorem}

\bigskip
\section  { THE TODA LATTICE}

\subsection*{ Definition of the  System}

The Toda lattice is a Hamiltonian system with Hamiltonian function 

\begin{equation}
  H(q_1, \dots, q_N, \,  p_1, \dots, p_N) = \sum_{i=1}^N \,  { 1 \over 2} \, p_i^2 +
\sum _{i=1}^{N-1} \,  e^{ q_i-q_{i+1}}  \ .
\end{equation} 
Hamilton's equations become

\begin{equation}
\begin{array}{lcl}
\dot q_j =p_j    \\
\dot p_j=e^{ q_{j-1}-q_j }- e^{q_j- q_{j+1}}  \ .
\end{array}
\end{equation}

\smallskip
\noindent
This system is  integrable. One can find a set of  independent functions  
$\{  H_1, \dots,  H_N \} $  which are constants of motion for Hamilton's equations.
To  determine the constants of motion, one uses Flaschka's transformation:
\begin{equation}
  a_i  = {1 \over 2} e^{ {1 \over 2} (q_i - q_{i+1} ) }  \ \ , \ \ \ \ \ \ \ \ \ \ \ \ \ \ 
             b_i  = -{ 1 \over 2} p_i  \ .     
\end{equation}

\smallskip
\noindent
Then

\begin{equation} 
\begin{array}{lcl}
 \dot a _i& = & a_i \,  (b_{i+1} -b_i )    \\  
   \dot b _i &= & 2 \, ( a_i^2 - a_{i-1}^2 ) \ . 
\end{array}
\end{equation}    
   
\smallskip
\noindent
These equations can be written as a Lax pair  $\dot L = [B, L] $, where $L$ is the Jacobi matrix 

\begin{equation}
 L= \pmatrix { b_1 &  a_1 & 0 & \cdots & \cdots & 0 \cr
                   a_1 & b_2 & a_2 & \cdots &    & \vdots \cr
                   0 & a_2 & b_3 & \ddots &  &  \cr
                   \vdots & & \ddots & \ddots & & \vdots \cr
                   \vdots & & & \ddots & \ddots & a_{N-1} \cr
                   0 & \cdots & & \cdots & a_{N-1} & b_N   \cr } \ , 
\end{equation} 

\smallskip
\noindent
and

\begin{equation}
    B =  \pmatrix { 0 & a_1 & 0 & \cdots & \cdots &  0 \cr
                 -a_1 & 0 & a_2 & \cdots & & \vdots  \cr
                    0  & -a_2 & 0 & \ddots &  & \cr
                    \vdots &  & \ddots & \ddots & \ddots & \vdots \cr
                     \vdots & & &  \ddots & \ddots & a_{N-1} \cr
                     0 & \cdots &\cdots &  & -a_{N-1}  & 0 \cr } \ .
\end{equation}

\bigskip
\noindent
 It follows that the  functions $ H_i={1 \over i} {\rm tr} \, L^i$ are  constants of
motion.

\smallskip
\noindent
{\it Remark:}  The Lax equation
\begin{equation}
\dot L (t) =[B(t), L(t)],  \ \ \ \ \ \ \ \ \ L(0)=L_0 
\end{equation}
can be solved by factorization.  First we perform  a Gram-Schmidt  factorization
  $e^{t L_0} =k(t) b(t)$, where $k(t)$ is 
orthogonal and $b(t)$ upper triangular. The solution is given by
\begin{equation}
L(t)=k(t)^{-1} \,  L_0 \,  k(t) \ .
\end{equation}
The form of the solution shows again that the functions ${\rm trace} \ L^k$ are 
independent of $t$.

\bigskip
\subsection*{Quadratic Toda bracket}

Consider ${\bf R}^{2N} $  with  coordinates $(q_1, \dots , q_N, p_1, \dots, p_N)$, the
standard symplectic bracket 
\begin{equation}
\{ f, g \}_s = \sum_{i=1}^N ( {\partial f \over \partial q_i} {\partial g 
\over \partial p_i} - {\partial f \over \partial p_i} { \partial g \over
\partial q_i} )   \ ,
\end{equation}

and the mapping $F: {\bf R}^{2N} \to {\bf R}^{2N-1}$ defined by

\begin{equation}
 F:  (q_1, \dots, q_N, p_1, \dots, p_N) \to (a_1,  \dots, a_{N-1}, b_1, \dots, b_N) \ .
\end{equation}

Define a bracket on ${\bf R}^{2N-1} $  by
\begin{equation}
\{f, g \} = \{ f \circ F, g \circ F \}_s \ .
\end{equation}

We obtain a bracket defined by
\begin{equation}
\begin{array}{lcl}
\{a_i, b_i \}& =&-a_i  \\
\{a_i, b_{i+1} \} &=& a_i    \ .
\end{array}
\end{equation}
All other brackets are zero. 
$H_1=b_1+b_2 + \dots +b_N$ is the only Casimir.   The Hamiltonian in this bracket is
 $H_2 = { 1 \over 2}\  { \rm tr}\  L^2$.  We later prove that $ \{  H_i, H_j \}=0$. 
  We   denote this bracket by $\pi_1$.   

\bigskip
The quadratic Toda bracket appears in conjunction with isospectral deformations of 
Jacobi matrices. First, let $\lambda $ be an eigenvalue of $L$ with normalized 
eigenvector $v$. Standard perturbation theory shows that

\begin{equation}
\nabla \lambda = (2 v_1 v_2, \dots, v_{N-1} v_N, v_1^2, \dots, v_N^2)^T :=U^{\lambda} \ ,
\end{equation}

where $\nabla  \lambda $ denotes $( { \partial \lambda \over \partial a_1}, \dots, {\partial \lambda
 \over \partial b_N})$. Some manipulations show that $U^{\lambda}$ satisfies 

\begin{equation}
M U^{\lambda} = \lambda N U^{\lambda} ,
\end{equation}
where $M$ and $N$ are skew-symmetric matrices. It turns out that $N$ is the matrix of 
coefficients of the Poisson tensor (67), and $M$, whose coefficients are quadratic functions
 of the $a$'s and $b$'s, can be used to define a new Poisson tensor. We denote this 
quadratic bracket by $\pi_2$. The defining relations are:

\begin{equation}
\begin{array}{lcl}
\{a_i, a_{i+1} \}&=&{ 1 \over 2} a_i a_{i+1} \\
\{a_i, b_i \} &=& -a_i b_i                    \\
\{a_i, b_{i+1} \}&=& a_i b_{i+1}    \\
\{b_i, b_{i+1} \}&=& 2\, a_i^2   \ ;
\end{array}
\end{equation}
all other brackets are zero.  This bracket has ${\rm det} \, L$ as Casimir and $H_1 =
{\rm tr}\, L$ is the Hamiltonian. The eigenvalues of $L$ are still in involution.
Furthermore, $\pi_2$ is compatible with $\pi_1$. We also have 
\begin{equation}
M \nabla \, \lambda_j= \lambda_j N \nabla \, \lambda_j  \  \ \ \ \ \ \ \ \ \forall j \ .
\end{equation}
So,
\begin{equation}
\begin{array}{rcl}
M \nabla { 1 \over l} \sum \lambda_j^l& =& \sum \lambda_j^{l-1} M \nabla \lambda_j  \\
 &&  \\
                                      &=& \sum \lambda_j^l N \nabla \lambda_j  \\ 
 & &     \\
                                      &=& N \nabla { 1 \over l+1} \sum \lambda_j^{l+1} \ .
\end{array}
\end{equation}
Therefore,
\begin{equation}
M \nabla H_l = N \nabla H_{l+1}  \ .
\end{equation}
These relations are similar to the Lenard relations for  the KdV equation.
 We will generalize them later.

\smallskip
Finally, we remark that further manipulations with the Lenard relations for the infinite
 Toda lattice,  followed by setting all but  finitely many $a_i$, $b_i$ equal to zero,
yield another Poisson bracket, $\pi_3$, which is cubic in the coordinates. The defining
relations for $\pi_3$ are:
\begin{equation}
\begin{array}{lcl}
\{a_i, a_{i+1} \}&=&a_i a_{i+1} b_{i+1}   \\   
\{a_i, b_i \}&=&     -a_i b_i^2-a_i^3 \\
\{a_i, b_{i+1} &=&     a_i b_{i+1}^2 +a_i^3 \\
\{a_i, b_{i+2} \}&=&     a_i a_{i+1}^2 \\
\{a_{i+1}, b_i \}&=&     -a_i^2 a_{i+1} \\
\{b_i, b_{i+1} \} &=&     2\, a_i^2 \, (b_i+b_{i+1}) \ ; 
\end{array}
\end{equation}
all other brackets are zero. The bracket $\pi_3$ is compatible with both $\pi_1$ and
$\pi_2$ and the eigenvalues of $L$ are still in involution. 
\smallskip

\bigskip

\subsection*{  Construction of master symmetries}

In  \cite{damianou1}  a sequence of master symmetries  $X_n$, for $n \ge -1$ was constructed. These
vector fields generate  
an infinite sequence of contravariant 2-tensors $\pi_n$, for $n\ge 1$. Before stating the theorem, we define an
equivalence relation on  the space of  Poisson tensors. A bracket 
is {\it trivial } if $H_i$ is a Casimir $\forall \, i$.      Two brackets 
are considered equivalent if their difference is a trivial bracket.  An example of a trivial
bracket is $\chi_i \wedge \chi_j$.  There are examples  of trivial brackets that are not of 
this form. We summarize the properties of $X_n$ and $\pi_n$ in the following:

\begin{theorem} 

\hfill \\
\smallskip
\noindent
{\it i) } $\pi_n$ are all Poisson.

\smallskip
\noindent
{\it ii) } The functions $H_n= { 1 \over n} {\rm Tr} \ L^n$ are in involution
 with respect to all of the $\pi_n$.

 \smallskip
 \noindent
 {\it iii)}  $X_n (H_m) =(n+m) H_{n+m} $.

 \smallskip
 \noindent
{\it iv)} $L_{X_n} \pi_m =(m-n-2) \pi_{n+m} $,  up to equivalence.

\smallskip
\noindent
{\it v)} $[X_n, \chi_l] =(l-1) \chi_{l+n} $, where $\chi _l$ is the 
Hamiltonian vector field generated by $H_l$ with respect to $\pi_1$. 

\smallskip
\noindent
{\it vi)} $\pi_n\  \nabla \  H_l =\pi_{n-1}\   \nabla \  H_{l+1} $,   
where $\pi_n$ denotes  the Poisson matrix  of the tensor $\pi_n$. 
\end{theorem}
\smallskip
We give an outline of the construction of the vector fields $X_n$. We
define $X_{-1} $ to be

\begin{equation}
 {\rm grad}\  H_1 = {\rm grad } \ {\rm Tr}\ L = \sum_{i=1}^N {\partial \ \over \partial b_i}  
\end{equation} 
 
\smallskip
\noindent
and $X_0$ to be the Euler vector field 

\begin{equation} 
 \sum _{i=1}^{N-1} a_i {\partial \ \over \partial a_i} + 
\sum_{i=1}^N b_i {\partial \ \over \partial b_i} \ .
\end{equation} 

\smallskip
\noindent
We want $X_1$ to satisfy 

\begin{equation}
       X_1 ({\rm Tr}\  L^n) =n {\rm Tr}\  L^{n+1} \ .
\end{equation} 

\smallskip
\noindent
One way to find such a vector field is by considering the equation

\begin{equation}
\dot L = [B, L] +L^2  \ .
\end{equation} 

\smallskip
\noindent
Note that the left hand side of this equation is a tridiagonal matrix while the 
right hand side is pentadiagonal. We look for $B$ as a tridiagonal matrix 

\begin{equation}
    B=   \pmatrix { \gamma_1 & \beta_1 & 0 & \cdots & \cdots \cr
                \alpha_1 & \gamma_2 & \beta_2 & \cdots & \cdots \cr
                0 & \alpha_2 & \gamma _3 & \beta _3 & \cdots \cr
                \vdots & \vdots & \ddots & \ddots & \ddots \cr}  \ . 
\end{equation}

\smallskip
\noindent 
 We want to choose the $\alpha_i$, $\beta _i$ and $\gamma_i$ so that the  
 right hand side of equation  (78) becomes tridiagonal. One  simple solution
 is \  $\alpha _n = -(n+1) a_n $, \ $\beta _n = (n+1) a_n $,\  $ \gamma_n =0$. 
  The vector field $X_1$  is defined by the right hand side of (78) and :

\begin{equation}
 X_1 = \sum_{n=1}^{N-1} \dot a_n {\partial \ \over \partial a_n} +
   \sum _{n=1}^N \dot b_n {\partial \ \over \partial b_n} \ ,
 \end{equation} 

  \smallskip
  \noindent
  where 

  \smallskip
  \noindent

\begin{equation}
 \dot a_n =  - na_n b_n + (n+2) a_n b_{n+1 }
\end{equation} 
  
\begin{equation}
\dot b_n = (2n+3) a_n^2 + (1-2n) a_{n-1}^2 + b_n^2  \ .
\end{equation}

\smallskip

  The construction of  the vector field $X_2$ is similar. We consider the equation
\begin{equation}
 \dot L =[B, L]+L^3 \ .
 \end{equation} 

 \smallskip
  \noindent
  The calculations are similar to those for $X_1$. The matrix $B$ is now
  pentadiagonal and the system of equations slightly more complicated.

\smallskip
 We
continue the sequence of master symmetries for $n \ge 3$ by

\begin{equation}
  [X_1, X_{n-1} ] = (n-2) X_n \ .
\end{equation}

\bigskip
\subsection*{Properties of $X_n$ and $\pi_n$.}

\bigskip

It is well known that $\pi_1$, $\pi_2$, $\pi_3$ satisfy Lenard 
relations
\begin{equation}
\pi_n \nabla H_l = \pi_{n-1} \nabla H_{l+1} \ , \ \ \ \ n=2,3 \ \ \ \ \forall l \ . 
\end{equation}
We want to show that these relations hold for all values of $n$. We denote
the Hamiltonian vector field of $H_l$ with respect to the $n$th bracket 
by $\chi_l^n$. In other words,
\begin{equation}
\chi_l^n=[\pi_n, H_l] \ .
\end{equation}
We prove the Lenard relations in an equivalent form

\bigskip
\begin{proposition}
$\chi_l^{n+1}=\chi_{l+1}^n \ . $
\end{proposition}

\smallskip
\noindent
{\bf Proof:}  To prove this we need the identity 
\begin{equation}
[X_1, \chi_l^n]=(n-3) \chi_l^{n+1} + (l+1) \chi_{l+1}^n 
\end{equation}
which follows easily from $X_1(H_l)=(l+1) H_{l+1}$ and (86).
Therefore,

\begin{equation}
\begin{array}{lcl}
(n-3) \chi_l^{n+1}& =& [X_1, \chi_l^n]- (l+1) \chi_{l+1}^n \\ 
                 &=& [X_1, \chi_{l+1}^{n-1} -(l+1) \chi_{l+1}^n  \\
                 &=& (n-4) \chi_{l+1}^n +(l+2) \chi_{l+2}^{n-1} -(l+1) 
                       \chi_{l+1}^n \\
                &=&(n-4) \chi_{l+1}^n+(l+2)\chi_{l+1}^n - (l+1) \chi_{l+1}^n \\
                 &=& (n-3) \chi_{l+1}^n    \ \ \ \ \ \ \ \ \ \    \Box   \ .

\end{array}
\end{equation}

\smallskip
Using the Lenard relations we can show that the functions $H_n$ are
in involution with respect to all of the brackets $\pi_n$.

\smallskip
\begin{proposition}
$\{H_i, H_j \}_n =0$, where $\{ \ ,\ \}_n$ is the bracket corresponding to  
 $\pi_n$. 
\end{proposition}

\smallskip
\noindent
{\bf Proof:} First we consider the Lie-Poisson Toda bracket. We have

\begin{equation}
\{H_1, H_j \}=0  \ \ \ \ \ \  \forall j \ ,
\end{equation}
since $H_1$ is a Casimir for $\pi_1$. Suppose that $\{ H_{i-1} , H_j\}=0$ 
 $\forall \ j$. 

\begin{equation}
\begin{array} {lcl}
i\{ H_i, H_j \} &=& \{ X_1(H_{i-1}), H_j \}  \\
        &=& -[\chi_j^1, [X_1, H_{i-1}]]      \\
       &=&[X_1, \{ H_{i-1}, H_j \}]+[H_{i-1}, (j+1) \chi_{j+1}^1 ] \\
     &=& (j+1) \{ H_{i-1} , H_{j+1} \} \\
     &=& 0    \ .
\end{array}
\end{equation}
Now we use induction on $n$. Suppose
\begin{equation}
\{ H_i, H_j \}_n =0 \ \ \ \  \forall i,\, j \ .
\end{equation}

\begin{equation}
\begin{array} {lcl}
\{H_i, H_j \}_{n+1} &=& \chi_i^{n+1}(H_j)  \\
              &=& \chi_{i+1}^n (H_j)  \\
               &=& \{ H_{i+1}, H_j \}_n  \\
             &=& 0   \ . \ \ \ \ \ \  \Box  
\end{array}
\end{equation}

\smallskip
It is straightforward to verify that the mapping 
\begin{equation}
f(a_1, \dots, a_{N-1}, b_1, \dots, b_N) =(a_1, \dots, a_{N-1}, 
1+b_1, \dots, 1+b_N)
\end{equation}
is a Poisson map between $\pi_2$ and $\pi_1+\pi_2$. Since $f$ is a 
diffeomorphism, we have the isomorphism 
\begin{equation}
\pi_2 \cong \pi_1 + \pi_2 \ .
\end{equation}
In other words, the tensor $\pi_2$ encodes sufficient information for 
both the linear and quadratic Toda brackets. An easy induction generalizes
 this result: i.e.,  
\begin{proposition}
\begin{equation}
\pi_n \cong \sum_{j=0}^{n-1}  \pmatrix{ n-1 \cr
                                         j }  \pi_{n-j}  \ .
 \end{equation}
\end{proposition}

\smallskip
The function $\mbox{ tr } \, L^{2-n}$, which is well-defined on the open
set $ \mbox{ det}\ L \not= 0$, is a Casimir for $\pi_n$, for $n \ge 3$. 
The proof uses the Lenard type relation

\begin{equation}
\pi_n \nabla \lambda = \lambda \pi_{n-1} \nabla \lambda 
\end{equation}
satisfied by the eigenvalues of $L$. To prove the last equation, one
uses the relation
\begin{equation}
\pi_n \sum \lambda_k^{l-1} \nabla \lambda_k = \pi_{n-1} \sum \lambda_k^l 
\nabla \lambda_k \ .
\end{equation}
But
\begin{equation}
\sum \lambda_k^{l-1} (\pi_k \nabla \lambda_k -\lambda_k \pi_{n-1} \nabla 
\lambda_k)=0 \ ,
\end{equation}
for $l=1,2, \dots, N+1$, has only the trivial solution because the
 (Vandermonde) coefficient determinant is nonzero. 

\bigskip
\begin{proposition}
For $n >2$, $\mbox{tr} \, L^{2-n}$ is a Casimir for $\pi_n$ on the open dense
set $\mbox{ det} \, L \not= 0$. 
\end{proposition}

\smallskip
\noindent
{\bf Proof:}  For $n=3$,

\begin{equation}
\begin{array}{lcl}
\pi_3 \nabla \mbox{tr}\, L^{-1} &=& \pi_3 \sum_k -{ 1 \over \lambda_k^2} \nabla \lambda_k \\
                    &=& \sum_k  -{ 1 \over \lambda_k^2} \lambda_k \pi_2 \nabla \lambda_k \\
                    &=& - \sum_k \pi_1 \nabla \lambda_k   \\
                     &=& - \pi_1 \nabla \mbox{ tr}\, L =\chi_1^1 =0 \ .
\end{array}
\end{equation}
For $n>3$ the induction step is as follows:

\begin{equation}
\begin{array}{lcl}
\pi_n \nabla \mbox{tr}\, L^{2-n} &=& \pi_n \nabla \sum_k { 1 \over \lambda_k^{n-2}}  \\
                                 &=& \sum_k (2-n) { 1 \lambda_k^{n-1}} \pi_n \nabla \lambda_k \\
                                 &=& \sum_k (2-n) { 1 \over \lambda_k^{n-1} } \lambda_k \pi_{n-1} \nabla \lambda_k \\
                                 &=&  { n-2 \over n-3} \pi_{n-1}  \nabla \mbox{tr}\, L^{3-n}  \\
                                 &=& 0 \ .  \Box 
\end{array}
\end{equation}

 \bigskip
  \subsection*{ Symmetries of Toda equations}

  In this subsection we find an  infinite sequence of evolution vector 
   fields that are symmetries of Toda equations (59).  
 A symmetry group of a system of differential equations 
 is  a Lie group acting on the space of independent and dependent
variables in such a way that solutions are mapped into other solutions. 
Knowing the symmetry  group allows one to determine some special types of solutions
 invariant under a subgroup of the full symmetry group, and in some cases
  one can  solve the equations completely.  
 The symmetry approach to solving differential equations can 
be found, for example, in the books of Olver \cite{olver2},   Bluman and Cole \cite{bluman1}, Bluman
 and Kumei \cite{bluman2}, and  Ovsiannikov \cite{ovsiannikov}. 
   \smallskip
   
  We  begin by writing  equations (59) in the form 
  \begin{equation}
\begin{array}{lcl}
 \Gamma _j& = &\dot a_j -a_j b_{j+1} +a_j b_j =0   \\
\Delta _j& =& \dot b_j -2 a_j^2 +2 a_{j-1}^2  =0  \ .
\end{array}
\end{equation}
 
\smallskip
\noindent
We look for symmetries of  Toda equations. i.e.,   vector fields of the 
form 
\begin{equation}
{\bf v} = \tau {\partial \over \partial t} + \sum_{j=1}^{N-1} \phi_j 
{\partial   \over \partial a_j} + \sum_{j=1}^N \psi_j {\partial \over \partial b_j }  \ ,  
\end{equation}
that  generate the symmetry group of the Toda System. 
\smallskip
\noindent
The first prolongation of ${\bf v}$ is 
\begin{equation}
 {\rm pr}^{(1)} {\bf v} = {\bf v} + \sum_{j=1}^{N-1} f_j { \partial \over 
\partial \dot a_j } + \sum_{j=1}^N g_j {\partial \over \partial \dot b_j} \ ,
\end{equation}

\smallskip
\noindent
where 
\begin{equation}
\begin{array}{lcl}
f_j&= &\dot{\phi}_j -\dot {\tau} \dot a_j  \\
g_j& = &\dot{\psi}_j -\dot {\tau} \dot b_j  \ . 
\end{array}
\end{equation}

\smallskip
\noindent
 The infinitesimal condition for a group to be a symmetry of the system is
\begin{equation}
\begin{array}{lcl}
{\rm pr}^{(1)} (\Gamma_j)  & =& 0 \\ 
 {\rm pr}^{(1)} (\Delta_j) & =&0  \ .  
\end{array}
\end{equation}
 
\smallskip
\noindent
Therefore we obtain the equations 
\begin{equation}
 \dot{\phi}_j - \dot{\tau} a_j (b_{j+1} -b_j) + \phi_j (b_j-b_{j+1} ) +
a_j \psi_j -a_j \psi_{j+1} =0   
\end{equation}

\begin{equation}
 \dot {\psi}_j - 2 \dot {\tau} (a_j^2 -a_{j-1}^2) -4 a_j \phi_j + 4 a_{j-1} \phi_{j-1} =0 \ .
\end{equation} 

\smallskip
\noindent
We first give some obvious solutions :

\smallskip
\noindent
{\it i)} \ \ $\tau  =0$,\  $ \phi_j =0 $,\  $\psi _j =1 $.
\bigskip
This is the vector field $X_{-1}$. 

\smallskip
\noindent
{\it ii)} \ \ $\tau =-1$, \ $\phi_j=0$, \ $ \psi_j=0$. 
\bigskip
The resulting vector field  is the time translation
$-{\partial \over \partial t} $ whose evolutionary representative is
\begin{equation}
 \sum_{j=1}^{N-1} \dot a_j {\partial \over \partial a_j} + \sum_{j=1}^N 
\dot b_j {\partial \over \partial b_j} \ . 
\end{equation}

\smallskip
\noindent
This is the Hamiltonian vector field $\chi_{H_2} $. It generates a Hamiltonian 
symmetry group.

\smallskip
\noindent
{\it iii)} \ \ $\tau = -t$, \ $ \phi_j=a_j$, \ $ \psi_j= b_j$. 
\bigskip
Then

\begin{equation}
{\bf v} = - t {\partial \over \partial t} + \sum_{j=1}^{N-1} a_j {\partial \over \partial  a_j}  
+ \sum_{j=1}^N b_j {\partial \over \partial b_j} =- t {\partial \over \partial t} + X_0 \ . 
\end{equation}

\smallskip
\noindent
This vector field generates the same symmetry as the evolutionary vector field 
\begin{equation}
X_0 + t \chi_{H_2}  \ .
\end{equation}

\smallskip
 We next look for some non-obvious solutions. The vector field $X_1$ is not
 a symmetry, so we add a term which depends on time. We try
\begin{equation}
\begin{array}{lcl}
   \phi_j &=& -j a_j b_j +(j+2) a_j b_{j+1} + 
                    t (a_j a_{j+1}^2 +a_j b_{j+1}^2-a_{j-1}^2 a_j -a_j b_j^2 )  \\
                    
 \psi _j &= &(2j+3) a_j^2 +(1-2j) a_{j-1}^2 +b_j^2 +  \\

         & &              + t (2 a_j^2 b_{j+1} +2 a_j^2 -2a_{j-1}^2 a_j -2a_{j-1}^2 b_j ) \ ,
\end{array}
\end{equation}
  \smallskip
  \noindent
  with  $\tau =0$.

 \smallskip
 \noindent
 A tedious but straightforward calculation shows that $\phi_j$, $\psi_j$ 
 satisfy (106) and (107). It is also straightforward to check that the vector field
 $\sum \phi_j {\partial \over \partial a_j} +\sum \psi_j {\partial \over \partial b_j} $
  is precisely  equal to $X_1 + t \chi_{H_3}$. The pattern suggests that 
  $X_n + t \chi_{H_{n+2} } $ is a symmetry of Toda equations.

 \bigskip
  \smallskip
  \noindent
  \begin{theorem} The vector fields $X_n +t \chi_{n+2} $ are symmetries of 
  Toda equations for $n\ge -1$. 
\end{theorem}

  \bigskip
  \noindent
  {\bf Proof :}  Note that $\chi_{H_1} =0$ because $H_1$ is a Casimir for the  
    Lie-Poisson  bracket.  We use the formula 
\begin{equation}
 [X_n, \chi_l] =(l-1) \chi_{n+l} \ .
\end{equation}

\smallskip
\noindent
In particular, for $l=2$, we have $[X_n, \chi_2] =\chi_{n+2} $. 

\bigskip
Since the Toda flow is Hamiltonian, generated by $\chi_2$, to
  show that $Y_n =X_n +t \chi_{n+2} $ are symmetries of Toda equations we must
verify the equation  
\begin{equation}
{\partial Y_n \over \partial t} + [\chi_2, Y_n] = 0 \ .
\end{equation} 

\smallskip
\noindent
But
\begin{equation}
\begin{array}{lcl}
{\partial Y_n \over \partial t} + [\chi_2,\  Y_n] &= &  {\partial Y_n \over
 \partial t} +[ \chi_2, \  X_n +t \chi_{n+2}   ]  \\ 
                                  & = &   \chi_{n+2} -[X_n,\  \chi_2] \\
                                  & = &\chi_{n+2} -\chi_{n+2} =0  \ . 
\end{array}           
\end{equation}
\hfill $\Box$

\subsection*{ A recursion operator for the Toda lattice}

\bigskip
One way of finding master symmetries of finite dimensional systems is the method
 used by some authors in the case of the finite nonperiodic Toda lattice. We briefly describe the procedure.

\smallskip
The first step is to define a second Poisson bracket on the space of canonical variables 
$(q_1, \dots, q_N, p_1,\dots, p_N)$. This bracket appears in Das and Okubo \cite{das} and 
Fernandes  \cite{fernandes}. We follow the notation from \cite{fernandes}. Let $J_0$ be the symplectic bracket  and 
define $J_1$  as follows:
\begin{equation}
\begin{array}{lcl}
\{q_i, q_j \}&=&  1   \\
\{p_i, q_i \}& =&p_i \\
\{ p_i, p_{i+1} \}&=& e^{q_i-q_{i+1}}   \ .
\end{array}
\end{equation}
Also define 
\begin{equation}
h_0 =  \sum_{i=1}^N p_i   \ \ \ \ \ \ \ \ \ \  \  h_1 =\sum_{i=1}^N {p_i^2 \over 2} +\sum_{i=1}^{N-1} e^{q_i-q_{i+1}}  \ .
\end{equation}
The recursion operator is defined by ${\cal R} =J_1 J_0^{-1}$.   It follows easily that the vector field 
\begin{equation}
Z_0 = \sum_{i=1}^N { N+1-2i \over 2} {\partial \over \partial q_i} +\sum_{i=1}^N p_i {\partial \over \partial p_i} 
\end{equation}
is a conformal symmetry for $J_0$, $J_1$ and $h_0$ and therefore, theorem 5 applies. The constants in Theorem 5 
 turn out to be $\lambda = -1$, $\mu =0$ and $\nu =1$.  We end up with the following deformation relations:

\begin{equation}
\begin{array}{lcl}
L_{Z_i}  J_j &=& (j-i-1) J_{i+j}  \\
             & &   \\

{[ Z_i, Z_j ]}  &=& (j-i) Z_{i+j}   \\
 & &  \\
{[Z_i, \chi_j]} &=& j \chi_{i+j}   
\end{array}
\end{equation}

 Recall the Flaschka transformation (58)  $F: {\bf R}^{2N} \to {\bf R}^{2N-1}$ defined by

\begin{equation}
 F:  (q_1, \dots, q_N, p_1, \dots, p_N) \to (a_1,  \dots, a_{N-1}, b_1, \dots, b_N) \ .
\end{equation}
The Poisson tensors $J_0$ and $J_1$ reduce to ${\bf R}^{2N-1} $.  They  reduce precisely
to the tensors $\pi_1$ and $\pi_2$ of section III. The mapping $F$ is a Poisson mapping 
between $J_0$ and $\pi_1$. It is also a Poisson mapping between $J_1$ and $\pi_2$.  The
 Hamiltonians $h_0$ and $h_1$ correspond to the reduced Hamiltonians $H_1$ and $H_2$ respectively.
The recursion operator ${\cal R}$ cannot be reduced. Actually, it is easy to see that there exists
no recursion operator in the reduced space.  The kernels of the
two Poisson structures $\pi_1$ and $\pi_2$ are different and, therefore, it is impossible to find
 an operator that maps one to the other.

\smallskip
The deformation relations (118) become precisely the deformation relations of Theorem 6. Of course, one
has to replace $j$ by $j-1$ in the formulas involving $J_j$ because of the difference in notation between 
\cite{damianou1} and \cite{fernandes}. We should note
that the vector field $Z_1$ corresponds to the vector $X_1$  up to addition of a Hamiltonian 
vector field. For this reason the reduced $J_i$ correspond to $\pi_i$.
 However $Z_2$ does not correspond to $X_2$.  This implies two things:

\smallskip
\noindent
{\it i)}  The master symmetry $X_2$ does not come from the given recursion operator. A question 
we could ask,  is whether every master symmetry in the reduced space comes from a recursion 
operator.

\smallskip
\noindent
{\it ii)} If one replaces $X_2$ by the reduced $Z_2$  then the relations in Theorem 6
 become exact and we do not need an equivalence relation.

\bigskip

\section  { RELATIVISTIC TODA SYSTEMS}

\bigskip
   In this section some results on the non-relativistic Toda lattice are extended to the 
case of the relativistic Toda Systems.  The main new result is the  hierarchy of
higher Poisson brackets which also exists in the finite nonperiodic 
Toda lattice. 

\smallskip

The relativistic Toda lattice was introduced and studied by Ruijsenaars in \cite{ruijsenaars}; see
also \cite{oevel3}, \cite{bruschi1}, \cite{bruschi2}.
In terms of canonical coordinates the relativistic Toda lattice is defined
by the Hamiltonian

\begin{equation}
 H(q_1, \dots, q_N, \,  p_1, \dots, p_N) = \sum_{j=1}^N  e^{p_j}
           f(q_{j-1} -q_j) f(q_j-q_{j+1})  \ , 
\end {equation}
\smallskip
\noindent
where  $f(x)=\sqrt{ 1+g^2 e^x}$ and, by convention, $q_0=-\infty$, 
$q_{N+1}=\infty$. The number $g$ is a coupling constant. To see the connection with
the non-relativistic Toda lattice one writes the equation in Newtonian form

\begin{equation}
\ddot q_{j} =g^2  \dot q_j( \dot q_{j-1} {  {\rm exp} (q_{j-1} -q_j) \over  1+g^2 {\rm exp} (q_{j-1} -q_j)   } -   
\dot q_{j+1} { {\rm exp}(q_j -q_{j+1})   \over 1+g^2  {\rm exp}(q_j -q_{j+1})  } ) \ ,
\end {equation}             
\smallskip
\noindent
$j=1,2, \dots, N$. Setting $\dot q_j =\dot Q_j +c$ and letting $c\to \infty$ and  $g\to 0$ in such a way that $gc=1$, one obtains
 the  equations of motion for 
 the classical Toda lattice
 \begin {equation}
 \ddot Q_j = e^{ Q_{j-1} -Q_j} - e^{ Q_j- Q_{j+1} }  \ .
 \end {equation}
\smallskip
\noindent
In the classical case  one uses a change of variables to
prove integrability. We follow 
the  same technique. 
Combining the changes of variables from   \cite{ruijsenaars}, \cite{bruschi1} and  \cite{bruschi2}, we set
\begin {equation}
\begin {array}{rcl}
a_j &=& g^2 e^{q_j -q_{j+1} +p_j}  f(q_{j-1} -q_j) /  f(q_j-q_{j+1})\\
b_j &=& \dot q_j -a_j 
\end {array}
\end {equation}
\smallskip
\noindent
In these variables the Hamiltonian is homogeneous quadratic and the
equations of motion become:

\begin {equation}
\begin {array}{rcl}
 \dot a _j & = & a_j (b_j-b_{j+1} +a_{j-1} -a_{j+1}  )   \\ 
   \dot b _j& = & b_j (a_{j-1} -a_j ) \ .  
\end{array}   
\end {equation}  
\smallskip
\noindent
These equations can be written as a Lax pair  $\dot L = [L, B] $, where $L$ is the
 matrix 
\begin{equation}
 L= \pmatrix { a_1+b_1 & a_1 & 0 & \cdots & \cdots & 0 \cr
                   a_2+b_2 &a_2+ b_2 & a_2 & 0 &\cdots    &0  \cr
                   a_3+b_3 & a_3+b_3 &a_3+ b_3 & a_3 &\cdots  & 0 \cr
                   \vdots & & \ddots & \ddots & & \vdots \cr
                   a_{N-1}+b_{N-1} &\cdots & & \ddots & a_{N-1}+b_{N-1} & a_{N-1} \cr
                   b_N & b_N &\cdots & \cdots &  & b_N   \cr } \ , 
\end{equation}
\smallskip
\noindent
and
\begin{equation}
B =  \pmatrix { 0 & a_1 & 0 & \cdots & \cdots &  0 \cr
                0 &-a_1 & a_2 & \cdots & & \vdots  \cr
                    0  & 0 & -a_2 & a_3 &\cdots  & \cr
                    \vdots &  & \ddots & \ddots & \ddots & \vdots \cr
                     \vdots & & &  \ddots & \ddots & a_{N-1} \cr
                     0 & \cdots &\cdots &  & 0& -a_{N-1}    \cr } \ .
\end{equation}
\smallskip
\noindent
This shows  that the  functions $H_n={1 \over n} {\rm Tr}\ L^n$ are
constants of motion and therefore the system is integrable.

\smallskip
 In the new coordinates $a_j$, $b_j$  the symplectic bracket  
 is transformed into a new quadratic Poisson bracket defined as follows:
\begin {equation}
\begin {array}{rl}
\{ a_i, a_{i+1} \} = & a_i a_{i+1}  \\
\{ a_i, b_i \}  =& -a_i b_i  \\
\{a_i, b_{i+1}  \} =& a_i b_{i+1}  \ .
\end {array}
\end {equation}
\smallskip
\noindent
All other brackets are zero. This bracket has ${\rm det}\, L = \prod_{i=1}^N b_i$
 as Casimir, and the eigenvalues of $L$ are in involution. Taking $H_1= {\rm Tr}\,
  L $ as Hamiltonian we obtain equations (124). We denote this bracket
 by $\pi_2$. Note that in the case of non-relativistic Toda lattice the 
symplectic bracket is transformed into a linear Poisson bracket.
  \smallskip

  We next define a linear bracket $\pi_1$  as follows: 
\begin {equation}
\begin{array} {rcl}
\{ a_i, b_i \} &=& -a_i \\
\{a_i, b_{i+1} \} &=& a_i \\
\{ b_i, b_{i+1} \} &=& -a_i  \ .
\end {array}
\end{equation}
\smallskip
\noindent
All other brackets are  zero. In this bracket ${\rm Tr}\, L$ is the 
Casimir and $H_2 = {1 \over 2} {\rm Tr}\, L^2 $  is the Hamiltonian. 
Therefore we have  a bi-Hamiltonian system, a situation similar to the
classical case.

\smallskip
To find higher Poisson brackets we work as in section III. We construct a vector
field $X_1$ which satisfies $X_1(H_m)=(m+1) H_{m+1}$. One possibility is

\begin{equation}
 X_1 = \sum_{i=1}^{N-1} r_i {\partial \ \over \partial a_i} +
   \sum _{i=1}^N  s_i {\partial \ \over \partial b_i} \ ,
   \end{equation}
  \smallskip
  \noindent
  where 
 
  \smallskip
  \noindent
  \begin{equation}
  \begin{array}{rcl}
 r_i &= & a_i^2 +(i+2) a_i b_{i+1} +(1-i) a_ib_i + (i+2) a_i a_{i+1} +(1-i) a_{i-1}a_i \\

s_i &=&  b_i^2+(i+1) a_i b_i +(1-i) a_{i-1} b_i  \ .
\end{array}
\end{equation}

Taking the Lie derivative of $\pi_2$ in the direction of $X_1$, we obtain
a cubic Poisson bracket $\pi_3$:
\begin{equation}
\begin{array}{rl}
\{a_i, a_{i+1} \}=& a_i^2 a_{i+1} +a_i a_{i+1}^2 +2 a_i a_{i+1} b_{i+1} \\
\{ a_i, a_{i+2} \} =& a_i a_{i+1} a_{i+2}  \\
\{ a_i, b_i \}=& -a_i b_i (a_i +b_i) \\
\{ a_i, b_{i+1} \} =&  a_i b_{i+1} (a_i +b_{i+1}) \\
\{ a_i, b_{i+2} \} =& a_i a_{i+1} b_{i+2}   \\
\{ a_{i+1}, b_i \} =& -a_i a_{i+1} b_i \\
\{ b_i, b_{i+1} \} =& a_i b_i b_{i+1} 
\end{array}
\end{equation} 

\smallskip
\noindent
This bracket is compatible with both $\pi_1$ and $\pi_2$ and the eigenvalues
of $L$ are still in involution.  The Casimir in this bracket is ${\rm tr} \,  L^
{-1}$.   Another iteration of the 
 procedure gives nothing new since $L_{X_1} \pi_3=0$.

In a similar way we construct a vector field $X_2$ which satisfies 
 $X_2 (H_m)=(m+2) H_{m+2} $. For $N=3$  we can take
 \begin {equation}
X_2 =\sum_{i=1}^2 r_i  {\partial \ \over \partial a_i }+
  \sum _{i=1}^3 s_i {\partial \ \over \partial b_i} 
  \end{equation}

  \smallskip
  \noindent
  where

  \begin{equation}
  \begin{array}{rcl}
r_1 & =& a_1\, (a_1^2 +5a_1 b_1  -a_2^2 + 2 a_2 b_1 - 2 a_2 b_2 -a_2 b_3 + 4 b_1^2 + 2 b_1 b_2 -b_2^2)  \\
r_2 &= & a_2  \, (3a_1^2+4a_1 a_2  + 3 a_1 b_1 + 6 a_1 b_2 + 2 a_1 b_3 + a_2^2+    \\  
    & \ &  4 a_2 b_2 + a_2 b_3 - 2 b_1 b_2 + 2 b_1 b_3 + 3 b_2^2 + 2 b_2 b_3 )  \\
s_1&=& b_1 \, (-2a_1^2- 2a_1 a_2 -a_1 b_1 -2 a_1 b_2 +b_1^2 ) \\
s_2&=& b_2 \, (3a_1^2+2 a_1 a_2 + 3a_1 b_1 +4 a_1 b_2  -a_2^2 + 2 a_2 b_1 -a_2 b_3 +b_2^2) \\
s_3&=& b_3\, (  2 a_2^2 + 2 a_1 a_2 - 2 a_2 b_1 + 2 a_2 b_2 + 3 a_2 b_3 +b_3^2 ) \ .

\end{array}
\end{equation}

\smallskip
\noindent
We define $\pi_4 = L_{X_2} \pi_2$. This bracket is Poisson and we still have 
involution of constants of motion.  Another iteration, $L_{X_2} \pi_4$, gives 
 a trivial bracket.
  \smallskip
  \noindent
The construction of $X_2$ allows  one to construct $X_3=[X_1, X_2]$ and, inductively, a sequence 
$X_1, X_2, \cdots $ which satisfies
\begin {equation}
[X_n, X_m]=(m-n) X_{n+m} \ . 
\end {equation}

\smallskip
\noindent
{\it Remark: } One should try to find  master symmetries of this system
 using the method of \cite{fernandes}. 
 In the case of the relativistic Toda equations we 
were unable to construct a second bracket which projects onto 
either the linear $(\pi_1 )$  or cubic $(\pi_3)$ bracket and 
therefore we cannot implement the same procedure.

\smallskip
\section{ GENERALIZED TODA SYSTEMS   ASSOCIATED WITH SIMPLE LIE GROUPS}
\vskip 2cm

\subsection*{ Definition of the systems}

In this section we consider mechanical systems which generalize the 
finite, nonperiodic Toda lattice. These systems correspond to Dynkin
diagrams. It is well known that irreducible root systems classify 
simple Lie groups. So, in this generalization for each simple Lie algebra
 there  exists a mechanical system of Toda type. 

\smallskip
The generalization is obtained from the following simple observation: In
terms of the natural basis $q_i$ of weights, the simple roots of $A_{n-1}$
are
\begin{equation}
q_1-q_2, q_2-q_3, \dots, q_{n-1}-q_n \ .
\end{equation}
On the other hand, the potential for the Toda lattice is of the form

\begin{equation}
e^{q_1-q_2} +e^{q_2-q_3} + \dots +e^{q_{n-1}- q_n}  \ .
\end{equation}
We note that the angle between $q_{i-1}-q_i$ and $q_i -q_{i+1}$ is 
$ { 2 \pi \over 3}$ and the lengths of $ q_i -q_{i+1} $ are all
equal. The Toda lattice corresponds to a Dynkin diagram of type
 $A_{n-1}$. 

\smallskip
More generally, we consider potentials of the form 
\begin{equation}
U=c_1 \,  e^{f_1(q)} + \dots + c_l \,  e^{f_l(q)}
\end{equation}
where $c_1, \dots, c_l$ are constants, $f_i (q)$ is  
linear  and $l$ is the
rank of the simple Lie algebra.  For each Dynkin diagram we 
construct a Hamiltonian system of Toda type. These systems 
are interesting not only because they are integrable, but also for their 
fundamental importance in  the theory of semisimple Lie groups. For example Kostant in
\cite{kostant}  shows that the integration of these systems and the theory 
of the finite dimensional representations of semisimple Lie
groups are equivalent.  

\smallskip
For reference, we give a complete list of the Hamiltonians for  
 each simple Lie algebra.

\bigskip
\noindent
$\underline { A_{n-1}}$

\medskip
$$ H= { 1 \over 2} \sum_1^n p_j^2 + e^{ q_1- q_2} + \cdots + e^{ q_{n-1}-q_n} $$

\medskip
\noindent
$\underline{ B_n}$
\bigskip
 $$ H= { 1 \over 2} \sum_1^n p_j^2 + e^{ q_1- q_2} + \cdots + e^{ q_{n-1}-q_n}
   + e^{q_n} $$
\medskip
\noindent
$\underline{ C_n}$
\medskip
 $$ H= { 1 \over 2} \sum_1^n p_j^2 + e^{ q_1- q_2} + \cdots + e^{ q_{n-1}-q_n}
   + e^{ 2 q_n} $$
\medskip
\noindent
$\underline{ D_n}$
\medskip
 $$ H= { 1 \over 2} \sum_1^n p_j^2 + e^{ q_1- q_2} + \cdots + e^{ q_{n-1}-q_n}
   + e^{q_{n-1} +q_n} $$
\medskip
\noindent
$\underline{ G_2}$
\medskip
 $$ H= { 1 \over 2} \sum_1^3 p_j^2 + e^{ q_1- q_2} +  e^{  -2 q_1 +q_2 +q_3}
    $$
\medskip
\noindent
$\underline{ F_4}$
\medskip
 $$ H= { 1 \over 2} \sum_1^4 p_j^2 + e^{ q_1- q_2} +e^{ q_2 -q_3} +e^{ q_3}
 + e^{ { 1 \over 2} (q_4 -q_1 -q_2 -q_3)} $$

\medskip
\noindent
$\underline{ E_6}$
\medskip
 $$ H= { 1 \over 2} \sum_1^8 p_j^2 +  \sum_1^4 e^{ q_j- q_{j+1} } +e^{ -(q_1+q_2)}
+ e^ { { 1 \over 2} ( -q_1 +q_2+ \dots + q_7-q_8)} $$

\medskip
\noindent
$\underline{ E_7}$
\medskip
 $$ H= { 1 \over 2} \sum_1^8 p_j^2 +  \sum_1^5 e^{ q_j- q_{j+1} } +e^{ -(q_1+q_2)}
+ e^ { { 1 \over 2} ( -q_1 +q_2+ \dots + q_7-q_8)} $$

\medskip
\noindent
$\underline{ E_8}$
\medskip
 $$ H= { 1 \over 2} \sum_1^8 p_j^2 +  \sum_1^6 e^{ q_j- q_{j+1} } +e^{ -(q_1+q_2)}
+ e^ { { 1 \over 2} ( -q_1 +q_2+ \dots + q_7-q_8)} $$
\bigskip

We should note that the Hamiltonians in the  list are not unique. For example, the
$A_2$ Hamiltonian is 

\begin{equation}
H={ 1 \over 2} p_1^2 + { 1 \over 2} p_2^2 + { 1 \over 2} p_3^2 + e^{q_1-q_2} + e^{q_2-q_3} \ .
\end{equation}
An equivalent system is 

\begin{equation}
H(Q_i, P_i) = { 1 \over 2} P_1^2 + { 1 \over 2} P_2^2 + e^ { \sqrt { { 2 \over 3}} ( \sqrt{3} Q_1 +Q_2 )}
+ e^{ -2 \sqrt{ { 2 \over 3}} Q_2} \ .
\end{equation}
 The second Hamiltonian is obtained from the first by using the canonical
 transformation 

\begin{eqnarray}
 Q_1 & =& { \sqrt2 \over 4} (q_1+q_2 -2 q_3) \\
             Q_2&=& { \sqrt 6 \over 4} (q_2-q_1)   \\
             P_1&=& { 2 \over \sqrt2}  (p_1+p_2)  \\
             P_2&=& { 2 \over \sqrt 6} (p_2-p_1)   \ .
\end{eqnarray}  

\bigskip
Another example  is the following two systems, both corresponding  
to a Lie algebra of type $D_4$:
\begin{equation}
\sum_{i=1}^4 { p_i^2 \over 2} + e^{q_1} +e^{q_2} +e^{q_3} +e^{ { 1 \over 2} (q_4 -q_1 -q_2-q_3)} 
\end{equation}

\begin{equation} 
\sum_{i=1}^4 { p_i^2 \over 2} + e^{q_1-q_2} +e^{q_2-q_3} +e^{ q_3-q_4} +e^{q_3+q_4} \ .
\end{equation}

\bigskip
\subsection*{ A rational bracket for $B_n$-Toda }

\bigskip
Another way to describe these generalized Toda systems, is to give a Lax pair 
representation in each case. It can be shown that the equation $\dot L =[B,L]$
is  equivalent to the equations of motion generated by the Hamiltonian 
 $H_2 = { 1 \over 2} {\rm tr} \, L^2 $ on the orbit through $L$ of the coadjoint 
action of  $B_{-}$ (lower triangular group) on the dual of its Lie algebra, ${\cal B}_{-}^*$. The space ${\cal B}_{-}^*$ can
be identified with the set of symmetric matrices. This situation, which corresponds
 to $sl(n, {\bf C})=A_{n-1}$ can be generalized to other semisimple Lie algebras. We
use notation and definitions from Humphreys \cite{humphreys}. 

\smallskip
Let ${\cal G}$ be a semisimple Lie algebra, $\Phi$ a root system for ${\cal G}$, 
$\Delta = \{ \alpha_1, \dots , \alpha_l \} $ the simple roots, $H$ a Cartan 
subalgebra and ${\cal G}_{\alpha}$ the root space of $\alpha$. We denote by $x_{\alpha}$ a 
generator of ${\cal G}_{\alpha}$.  Define  
\begin{equation}
{\cal B}_{-}=H \oplus \sum_{\alpha <0} {\cal G}_{\alpha} \ .
\end{equation}
There is an automorphism $\sigma$ of ${\cal G}$, of order 2, satisfying 
$\sigma (x_{\alpha})=x_{- \alpha}$ and $\sigma ( x_{-\alpha})= x_{\alpha}$. Let ${\cal K}=
\{ x \in \, {\cal G} \, \vert \, \sigma (x)=-x  \}$. Then we have a direct sum 
decomposition ${\cal G}={\cal B}_{-} \oplus {\cal K}$.  The Toda flow is coadjoint flow on 
${\cal B}_{-}^*$ and the coadjoint invariant functions on ${\cal G}^*$, when restricted to 
${\cal B}_{-}^*$ are still in involution.
\smallskip
The Jacobi elements are of the form
\begin{equation}
L=\sum_{i=1}^l \, b_i h_i + \sum_{i=1}^l a_i \, (x_{\alpha_i} + x_{-\alpha_i}) \ .
\end{equation}
We define
\begin{equation}
B=\sum_{i=1}^l a_i \,  (x_{\alpha_i }- x_{-\alpha_i}) \ .
\end{equation}
The generalized Toda flow takes the Lax pair form:
\begin{equation}
\dot L=[B,L]  \ .
\end{equation}

\smallskip
The $B_n$ Toda systems were shown to be Bi-Hamiltonian. The second bracket 
 can be found in \cite{damianou2}. It turned out to be a rational bracket and it was obtained by
using Dirac's constrained bracket formula (30).  The idea is to use the inclusion of $B_n$ into $A_{2n}$  and to 
restrict the hierarchy of brackets from $A_{2n}$ to $B_n$ via Dirac's bracket.
Straightforward restriction does not work. We briefly describe the procedure in the
case of $B_2$. 

\bigskip
The Jacobi matrices  for $A_4$ and $B_2$ are given by:

\begin{equation}
 L_{A_4} =  \pmatrix { b_1 & a_1 & 0 & 0& 0 \cr
              a_1 & b_2 & a_2 & 0& 0 \cr
              0& a_2 & b_3 & a_3 & 0 \cr
              0 & 0& a_3 & b_4 & a_4  \cr
               0& 0&0& a_4 & b_5 }  \ ,
 \end{equation}

and
\begin{equation}
 L_{B_2}=  \pmatrix { b_1 & a_1 & 0 & 0& 0 \cr
              a_1 & b_2 & a_2 & 0& 0 \cr
              0& a_2 & b_3 & -a_2 & 0 \cr
              0 & 0& -a_2 &  2 b_3 -b_2 & -a_1  \cr
               0& 0&0& -a_1 & 2 b_3 -b_1  }  \ .
\end{equation}
 Note that $L_{A_4}$ lies in ${\rm gl} (4, {\bf C})$  instead of $ {\rm sl} (4, {\bf C})$. Therefore we have
added an additional variable in $L_{B_2}$. We define

\begin{eqnarray}
p_1 &=& a_1 +a_4  \nonumber \\
p_2& = & a_2 +a_3\nonumber \\
p_3&= & b_1 +b_5 - 2 b_3\nonumber \\
p_4 &=& b_2 +b_4 -2 b_3  \ .
\end{eqnarray}
It is clear that we obtain $B_2$ from $A_4$ by setting $p_i=0$ for
$i=1,2,3,4$. We calculate the matrix $P=\{ p_i, p_j \}$.  The bracket 
  used  is the quadratic Toda  (70) on $A_4$.

\begin{eqnarray}
\{p_1, p_2 \}&=& { 1 \over 2} (a_1 a_2 -a_3 a_4)\nonumber \\
\{p_1, p_3 \}&=& a_4 b_5 -a_1 b_1 \nonumber \\
\{p_1 ,p_4 \}&=& a_1 b_2 -a_4 b_4 \nonumber \\
\{p_2, p_3 \}&=& 2(a_3 b_3 -2 a_2 b_3 )\nonumber \\
\{p_2, p_4 \}&=& a_3 b_4 + 2 a_3 b_3 -2 a_2 b_3 -a_2 b_2\nonumber \\
\{p_3, p_4 \}&=& -2 a_4^2 -4 a_3^2 + 4 a_2^2 +2 a_1^2  \ .
\end{eqnarray}

If we evaluate at a point in $B_2$ we get

\begin{eqnarray}
\{p_1, p_2 \}&=& 0   \nonumber \\
\{p_1, p_3 \}&=& -2 a_1 b_3    \nonumber \\
\{p_1 ,p_4 \}&=& 2 a_1 b_3 \nonumber \\
\{p_2, p_3 \}&=&  -4 a_2 b_3   \nonumber \\
\{p_2, p_4 \}&=& -6 a_2 b_3    \nonumber \\
\{p_3, p_4 \}&=&  0  \ .
\end{eqnarray} 

Therefore the matrix $P$ is given by

\begin{equation}
P=\pmatrix { 0 &0& -2 a_1 b_3 & 2 a_1 b_3  \cr
             0 &0& -4 a_2 b_3 & -6 a_2 b_3 \cr
             2 a_1 b_3 & 4 a_2 b_3 & 0 & 0 \cr
             -2 a_1 b_3 & 6 a_2 b_3 & 0 & 0   \ , }
\end{equation}
and $P^{-1}$ is the matrix 

\begin{equation}
P^{-1}= \pmatrix  { 0 & 0 & { 3 \over 10 a_1 b_3 } & -{ 1 \over 5 a_1 b_3} \cr
          0&0 & { 1 \over 10 a_2 b_3 } & { 1 \over 10 a_2 b_3}  \cr
          -{  3 \over 10 a_1 b_3 } & { 1 \over 5 a_1 b_3} & 0 &0 \cr 
            -{ 1 \over 10 a_2 b_3} & -{ 1 \over 10 a_2 b_3 }& 0 & 0  \ . } 
\end{equation}
Using Dirac's formula we obtain a homogeneous quadratic bracket on $B_2$ given by:

\begin{eqnarray}
\{a_1, a_2 \} &=& { a_1 a_2 ( 3 b_3 -b_2-2 b_1) \over 10 b_3}\nonumber \\
\{a_1, b_1 \} &=& { -a_1 (10 b_1 b_3 -2 b_1 b_2 -3 b_1^2 -a_1^2) \over 10 b_3} \nonumber \\
\{ a_1, b_2 \}&=& { a_1 (10 b_2 b_3 -3 b_2^2 -2 b_1 b_2 -4 a_2^2 -a_1^2) \over 10 b_3}\nonumber \\
\{a_1 , b_3 \}&=& { a_1 (b_2-b_1) \over 5} \nonumber \\
\{a_2, b_1 \}&=& { a_2 ( 2 b_1 b_3 -2 b_1 b_2 +a_1^2 ) \over 10 b_3} \nonumber \\
\{a_2, b_2 \}&=& { -a_2 ( 8 b_2 b_3 -3 b_2^2 -6 a_2^2 -4 a_1^2 ) \over 10 b_3} \nonumber \\  
\{a_2, b_3 \}&=& {a_2 (b_3-b_2) \over 5 } \nonumber \\
\{ b_1,b_2 \} &=& { 10 a_1^2 b_3 -3 a_1^2 b_2 -2 a_2^2 b_1 -3 a_1^2 b_1  \over 5 b_3}\nonumber \\
\{ b_1,b_3 \} &=& { 2 a_1^2 \over 5} \nonumber \\
\{ b_2,b_3 \}&=& { 2 \over 5} (a_2^2 -a_1^2)      \ . 
\end{eqnarray}
The bracket satisfies the following properties which are analogous to the quadratic $A_n$ Toda (70).

\smallskip
\noindent 
{\it i)} It is a homogeneous quadratic Poisson bracket.

\smallskip
\noindent 
{\it ii)} It is compatible with the $B_2$ Lie-Poisson bracket.

\smallskip
\noindent 
{\it iii)} The functions $H_n = { 1 \over n} \, {\rm tr} \, L^n$ are in involution in this
bracket.

\smallskip
\noindent 
{\it iv)} We have Lenard type relations $\pi_2 \nabla H_i = \pi_1 \nabla H_{i+1} $ where $\pi_1$,
$\pi_2$ are the component matrices of the linear and quadratic $B_2$ Toda brackets respectively.

\smallskip
\noindent 
{\it v)} The function ${\rm det}\, L$ is the Casimir.

\smallskip
 We conjecture that the only bracket satisfying all five properties is the one just
obtained. In \cite{damianou2}  a quadratic bracket is defined which satisfies properties {\it i)- iv)} but 
not {\it v)}.

\bigskip

\subsection*{  A recursion operator for  $B_n$  Toda systems}

 In this section, we use  a different approach to  show that polynomial brackets exist in the case of $B_n$ Toda  systems. 
We will prove  that these systems  possess a recursion operator and we will
construct an infinite sequence of compatible Poisson brackets in which the constants of motion are
in involution.

\smallskip
 The Hamiltonian for $B_n$ is 

\begin{equation}
 H= { 1 \over 2} \sum_1^n p_j^2 + e^{ q_1- q_2} + \cdots + e^{ q_{n-1}-q_n}
   + e^{q_n}  \ .
\end{equation} 

 We make a  Flaschka-type  transformation 

\begin{equation}
  a_i  = {1 \over 2} \,  e^{ {1 \over 2} (q_i - q_{i+1} ) }   \ \ \ \ \ \ \ \ \ \  a_n= { 1 \over 2} e^{ { 1 \over 2} q_n }
\nonumber
\end{equation}

\begin{equation}
             b_i  = -{ 1 \over 2} p_i    \ .  
\end{equation}

These equations can be written as a Lax pair  $\dot L = [B, L] $, where $L$ is the
 matrix 
\begin{equation}
  \pmatrix { b_1 &  a_1 &  & &  &    &    & \cr
                   a_1 &  \ddots  & \ddots  &   & &&   &  \cr
                    & \ddots & \ddots& a_{n-1} &  & &   &  \cr
                   &  & a_{n-1} & b_n & a_n &  & &  \cr
                    & &  & a_n & 0 &  -a_n     & & \cr
                    & & & & -a_n & -b_n & \ddots &  \cr
                    &  &&&& \ddots & \ddots & -a_1  \cr
                       &&&&& & -a_1 & -b_1     \cr }   \ .
\end{equation}

\smallskip

In the new variables $a_i$, $b_i$ the symplectic bracket  $\pi_1 $ is given by

\bigskip
\noindent
\begin{equation}
\begin{array}{lcl}
\{ a_i, b_i \} & = &-a_i  \\
\{ a_i, b_{i+1} \}& =& a_i \ .
\end{array} 
\end{equation}

\smallskip

The invariant polynomials for $B_n$, which we denote by
\begin{equation}
  H_2, \  H_4, \  \dots \ H_{2n}
\end{equation}

are defined by $H_{2i} = { 1 \over 2i} \  { \rm Tr} \ L^{2i}  $.

\smallskip

We look for a bracket $\pi_3$ which satisfies
\begin{equation}
 \pi_3 \ \nabla \ H_2 = \pi_1 \ \nabla  \ H_4  \ .
\end{equation}

\smallskip
 Using trial and error,  we end up with the following homogeneous cubic bracket  $\pi_3$.

\begin{equation}
\begin{array}{lcl}
\{ a_i, a_{i+1} \}& = &a_i a_{i+1} b_{i+1}  \\
\{ a_i, b_i \}& = &-a_i b_i^2 -a_i^3   \ \ \ \ \ \ \ \ \ \ i=1,2, \dots , n-1 \\ 
\{ a_n,b_n \}& = &-a_n b_n^2 -2 a_n^3  \\ 
\{a_i, b_{i+2} \}& = &a_i a_{i+1}^2   \\ 
\{ a_i, b_{i+1} \}& =& a_i b_{i+1}^2 + a_i^3  \\ 
 \{ a_i, b_{i-1} \} &=& -a_{i-1}^2 a_i  \\ 
\{ b_i, b_{i+1} \}&= &2 a_i^2 (b_i +b_{i+1})  \ .
\end{array}
\end{equation}
\smallskip
We summarize the properties of this new bracket in the following:

\begin{theorem}  The bracket $\pi_3$ satisfies:

\smallskip
\noindent
1. $\pi_3 $ is Poisson

\smallskip
\noindent
2. $\pi_3$ is compatible with $\pi_1$.

\smallskip
\noindent
3. $H_{2i}$ are in involution.

\smallskip
\noindent
4. $\pi_{j+2} \  {\rm grad}\  H_{2i} = \pi_j \  {\rm grad} \  H_{2i+2} \ \ \ \ \ \forall \ i, \ j \   \ . $ 

\smallskip
Define $N=\pi_3 \pi_1^{-1}$. Then $N$ is a recursion operator.
 We obtain a hierarchy  
$$\pi_1, \pi_3, \pi_5, \dots $$
consisting of compatible Poisson brackets of odd degree in which   the constants of 
motion are in involution. 
\end{theorem}

\smallskip
\noindent
The proofs of {\it 1.} and { \it 2.} are straightforward verification of the Jacobi identity. Since (164) holds,
{\it  4.} follows from properties of the recursion operator.  {\it 3.} is a consequence of {\it 4.}

\bigskip

\section   { A TRI-HAMILTONIAN FORMULATION   OF THE FULL KOSTANT-TODA LATTICE}

 The  full Kostant-Toda lattice  is another variation of the Toda lattice. We briefly describe the system:
   In \cite{kostant} Kostant conjugates the matrix $L$ in (60)  by a diagonal matrix
  to obtain a matrix of the form

\begin{equation}
X= \pmatrix { b_1 &  1 & 0 & \cdots & \cdots & 0 \cr
                   a_1 & b_2 & 1 & \ddots &    & \vdots \cr
                   0 & a_2 & b_3 & \ddots &  &  \vdots \cr
                   \vdots & \ddots & \ddots & \ddots & & 0 \cr
                   \vdots & & & \ddots & \ddots & 1 \cr
                   0 & \cdots &  \cdots & 0 & a_{n-1} & b_n   \cr } \ . 
\end{equation}

The equations take the form 

\begin{equation}
\dot X(t)=[ X(t), P \, X(t)]
\end{equation}
where $P$ is the projection onto the strictly lower triangular 
part of $X(t)$.   
This form is convenient in applying Lie theoretic techniques to 
describe the system.

\smallskip
To obtain the full Kostant-Toda lattice we fill the lower triangular
 part of $X$ in (166) with  additional variables. ($P $ is again the projection
 onto the strictly lower part of $X(t)$).  So, using the notation from \cite{singer},   \cite{ercolani}, \cite{flaschka3}. 
\begin{equation}
\dot X(t)=[ X(t), P \, X(t)] \ ,
\end{equation}
where $X$ is in $ \epsilon + B_-$ and $ P \, X$ is in $N_-$.  
$B_-$ is the Lie algebra of lower triangular matrices and $N_-$ is 
the Lie algebra of strictly lower triangular matrices.
 In the case of $sl(4,{\bf C})$  the matrix $X$ 
 has   the form 
\begin{equation} 
X= \pmatrix { f_1 &  1 & 0 & 0 \cr
                   g_1 & f_2 & 1 &  0 \cr
                   h_1 & g_2 & f_3 & 1   \cr
                   k_1 & h_2 & g_3 & f_4   \cr } \ , 
\end{equation}
with $\sum_i f_i =0$.

\smallskip

 We now   apply the method of section III to generate  nonlinear Poisson brackets 
for the full Kostant-Toda lattice.
 The Poisson brackets are deformations of the 
Lie Poisson bracket on $B_+^* $ and they are obtained by using master symmetries.
  The main difference in this version of Toda lattice is that the
sequence of Poisson brackets is not infinite.  The first three tensors are 
Poisson, but the remaining ones fail to satisfy the Jacobi identity.
 We are therefore in the situation investigated in \cite{li} and 
  \cite{oevel4}. In fact, some of the results (and proofs)  of Li and Parmentier in  
  \cite{li},  on the 
full symmetric Toda  carry over to this system almost without change.  The connection
with R-matrices and the full symmetric Toda lattice  is worth further investigation.

\smallskip

The vector fields in the construction are unique up to addition of a Hamiltonian
 vector field. Similarly, the Poisson brackets are unique up to addition of a 
 trivial bracket.  By generating the second Hamiltonian structure, which turns out to be quadratic,
  we obtain  a bi-Hamiltonian system.  One can use this 
fact to prove involutivity of integrals as in Ratiu \cite{ratiu}.
Furthermore, a third Poisson structure is found, which leads to  a 
tri-Hamiltonian formulation of the equations.  In this system, 
all constants of motion, polynomial and rational,   are in involution with respect to all
 three of  the Poisson brackets.

\medskip

  Let ${\cal G}= sl(n)$, the Lie algebra of $n \times n$ matrices of trace zero.
 Using the decomposition  ${\cal G}= B_+  \oplus  N_-$  we can identify   $B_+^*$ with the 
annihilator of $N_-$  with respect to the trace form. This annihilator is $B_-$.  Thus we can
identify $B_+^*$ with $B_-$ and therefore with $\epsilon +B_-$ as well.   The Lie Poisson bracket in
 the case of $sl(4)$ is given by the  following defining relations:

\smallskip
\noindent
$\{ g_i, g_{i+1} \} = h_i $,

\noindent
$\{g_i, f_i \} = - g_i$,

\noindent 
$\{ g_i, f_{i+1} \}= g_i$,

\noindent 
$\{ h_i, f_i \} =-h_i $,

\noindent 
$\{  h_i, f_{i+2} \}= h_i$,

\noindent 
$\{ g_1, h_2 \} = k_1$,

\noindent 
$\{ g_3, h_1 \} =- k_1$,

\noindent 
$\{ k_1, f_1 \} = -k_1$,

\noindent 
$\{ k_1, f_4 \} =k_1  $.

\smallskip
\noindent
All other brackets are zero. Actually,  we calculated the brackets on $gl(4, {\bf C})$; the trace of $X$ now becomes a 
Casimir. The Hamiltonian in this bracket is $H_2 = { 1 \over 2}\  { \rm Tr}\  X^2$.

\smallskip
\noindent
{\it Remark:} If we use a more conventional notation for the matrix $X$, i.e., 
$x_{ij}$ for  $i \ge j$, $x_{ii+1}=1$, and all other entries zero, then the 
bracket is simply $ \{ x_{ij}, x_{kl} \} = \delta_{li} x_{kj} -\delta_{jk} 
x_{il} $.

\smallskip
 The functions  $H_i= { 1 \over i} \, {\rm Tr} \ X^i$  are still in involution but they are  not enough to
ensure integrability. There are, however,  additional integrals and the interesting
 feature of  this  system is that the additional integrals turn out to be rational
functions of the entries of $X$.  We describe the constants of motion following references
 \cite{singer},  \cite{ercolani}, \cite{flaschka3}.

\smallskip
For $k=0, \dots , [ { (n-1) \over 2}]$,\, denote by $( X- \lambda \, { \rm Id})_{ (k)}$ the 
result of removing the first $k$ rows and last $k$ columns from $X- \lambda \,{\rm  Id}$, 
and let 
\begin{equation}
{\rm det} \ ( X- \lambda \, { \rm Id})_{ (k)} = E_{0k} \lambda ^{n- 2k} + \dots + E_{n-2k,k} \ .
\end{equation}
Set 
\begin{equation}
{  {\rm det} \ ( X- \lambda \, { \rm Id})_{ (k)}  \over E_{0k}} = \lambda^{ n-2k} + I_{1k} \lambda ^
{n-2k-1} + \dots + I_{n-2k,k} \ .
\end{equation}
The functions $I_{rk}$, $r=1, \dots, n-2k$, are constants of motion for (168).

\smallskip
So, in the case of $gl(4, {\bf C})$ the additional integral is 

\begin{equation}
I_{21}= { g_1 g_2 g_3 - g_1 f_3 h_2 - f_2 g_3 h_1 + h_1 h_2  \over k_1} + f_2 f_3 -g_2 \ ,
\end{equation}

\noindent
and 
\begin{equation}
I_{11}={ g_1 h_2 + g_3 h_1  \over k_1}-f_2 -f_3 
\end{equation}
 is a Casimir.

\bigskip
 We want to define a second bracket $\pi_2$ so that $H_1$ is the Hamiltonian and

\begin{equation}
\pi_2 \ {\rm grad} \ H_1 = \pi_1 \ {\rm grad} \ H_2  \ .
\end{equation}

\noindent
i.e.,  we want to construct a bi-Hamiltonian pair. We will achieve  this by 
finding a master symmetry.

\smallskip
 To construct $X_1$, we  consider the equation

\begin{equation}
\dot{X}=[ Y,X]+ X^2 \ .
\end{equation}

 $Y$ is chosen in such a way  that the equation is
consistent. One solution is 
\begin{equation}
Y= \sum_{i=1}^n \alpha_i E_{ii} +\sum_{i=1}^{n-1} \beta_i E_{i,i+1}  \ ,
\end{equation}
where

\smallskip
\noindent
 $\beta_i=i$,

\smallskip
\noindent
 $\alpha_i= i f_i + \sum_{k=1}^{i-1} f_k$.

\smallskip
The vector field $X_1$ is defined by the right hand side of (175).
For example, in $gl(4, {\bf C})$  the components of $X_1$ are:  

\medskip
\noindent
 $ X_1(f_1)= 2 g_1+f_1^2 $,

\noindent
$ X_1 (f_2)=3 g_2+f_2^2$,

\noindent 
$X_1(f_3)=  -g_2 +4 g_3 +f_3^2 $,

\noindent
$X_1 (f_4)=  -2 g_3+f_4^2 $,

\noindent
$ X_1 (g_1)=  3 h_1 + g_1 f_1 + 3 g_1 f_2 $,

\noindent
$X_1(g_2)=  4 h_2+ 4 g_2 f_3 $,

\noindent
$ X_1(g_3)= -h_2 -g_3 f_3 + 5 g_3 f_4$,

\noindent
$ X_1 (h_1)=g_1 g_2+ 4 k_1 +h_1 f_1 +h_1 f_2 +4 h_1 f_3$,

\noindent
$ X_1 (h_2)= g_2 g_3+h_2 f_3 +5 h_2 f_4 $,

\noindent
$X_1 (k_1)= g_3 h_1 +g_1 h_2+k_1 f_1 +k_1 f_2 +k_1 f_3+ 5 k_1 f_4$.

\smallskip
The second bracket, $\pi_2$,  is defined by taking the Lie derivative of $\pi_1$ in 
the direction of $X_1$.

\smallskip
 Similarly, we define $\pi_3 = L_{X_1} \pi_2$. Another iteration of the 
 procedure gives nothing new since $L_{X_1} \pi_3=0$.

\smallskip
\noindent
We  define  $X_0$ to be the Euler vector field and  $X_{-1} = {\rm grad}
  \ H_1$.

\smallskip
To construct the vector field $X_2$ we consider the equation

\begin{equation}
\dot{X}=[ Y,X]+ X^3
\end{equation}

\noindent
We take 
\begin{equation}
Y=\sum_{i=1}^n \alpha_i E_{ii} + \sum_{i=1}^{n-1} \beta_i E_{i, i+1} + \sum_{i=1}^
{n-2} \gamma_i E_{i, i+2} 
\end{equation}
with

\medskip
\noindent
$\alpha_i = i f_i^2 + \sum_{k=1}^{i-1} f_k^2 +f_i \sum_{k=1}^{i-1} f_k
+i[g_i+ g_{i-1} ] + 2 \sum_{k=1}^{i-2} g_k $,

\medskip
\noindent
$\beta_i = i [ f_{i+1} +f_i ] + \sum_{k=1}^{i-1}f_k   $,

\medskip
\noindent
$\gamma_i =i $.

In the formulas we take $g_i =0 $ for $ i = 0, n  $.  

\smallskip
 We complete the sequence of the 
master symmetries  $X_i$ for $i \ge 3$ by taking Lie brackets. 
 We summarize the results in a Theorem. The proofs are almost 
identical with similar ones for the classical Toda lattice.

\smallskip

\begin{theorem}

 There exists  a sequence of vector fields $X_i$, for $i \ge -1$, and a sequence of
  contravariant 2-tensors $\pi_i$, $i\ge 1$, 
satisfying :

\smallskip
\noindent
{\it i) } $\pi_i$ are  Poisson for $i=1,2,3$.

\smallskip
\noindent
{\it ii) } The functions $H_i= { 1 \over i} {\rm Tr} \ X^i$ are in involution
 with respect to all of the $\pi_i$. Actually all invariants, including the 
rational ones, are in involution.

 \smallskip
 \noindent
 {\it iii)}  $X_i (H_j) =(i+j) H_{i+j} $.

 \smallskip
 \noindent
{\it iv)} $L_{X_i} \pi_j =(j-i-2) \pi_{i+j} $, up to equivalence.

\smallskip
\noindent
{\it v)} $[X_i, \chi_l] =(l-1) \chi_{l+i} $. 

\smallskip
\noindent
{\it vi)} $\pi_i\ {\rm grad }\  H_l =\pi_{i-1}\  {\rm grad}\  H_{l+1} $, $i=2,3$.

\smallskip
\noindent
{\it vii)}  A polynomial  Casimir for $\pi_2$ is $ {\rm det}\, X$ and 
  $ {\rm Tr}\ X^{-1} $ is a Casimir for $\pi_3$.  

\end{theorem}
\bigskip
\noindent
{\it Remark 1)} \ The master symmetries $X_i$ preserve  constants of 
motion. It is interesting to see where the rational
invariants are mapped in a low dimensional example. We consider the
case of $gl(5, {\bf C})$.  We have four rational invariants denoted
by $K_1=I_{11}$, $K_2=I_{21}$, $K_3=I_{31}$ and $K_4=I_{12}$.
The master
 symmetry $X_1$ behaves in the following way:

\smallskip
\noindent
{\it i)} $X_1(K_1)= 2 K_2 +K_1^2$

\smallskip
\noindent
{\it ii)} $X_1(K_2)= 3 K_3 + K_1 K_2 $

\smallskip
\noindent
{\it iii)} $X_1(K_3)= K_1 K_3 $

\smallskip
\noindent
{\it iv)} $X_1(K_4)= K_4^2$.

\smallskip
\noindent
The master symmetry $X_2$ acts in a more complicated
way. For example,

\noindent
 $X_2(K_3)=   { 1 \over 120} H_1^5 - { 1 \over  6} H_1^3 H_2 +  H_1 K_1 K_3
- { 1 \over 2} H_1^2 K_3 + { 1 \over 2} H_1 H_2^2  + { 1 \over 2} H_1^2 H_3 - H_1 H_4 + K_2 K_3 - H_2 H_3 
+K_3 H_2 +H_5$\,  !

\bigskip
\noindent
{\it Remark 2)}  We also have Lenard-type relations for the rational 
invariants. For example in the case of  $gl(5, {\bf C})$

\smallskip
\noindent
{\it i)} $ \pi_1 { \rm grad} \,  K_2 = \pi_2 {\rm grad} \, K_1 $

\smallskip
\noindent
{\it ii)} $ \pi_1 { \rm grad} \,  K_3 = \pi_2 {\rm grad} \, K_2 $

\smallskip
\noindent
{\it iii)} $ \pi_1 { \rm grad} \,  K_4 = \pi_2 {\rm grad} \, K_3 $

\smallskip
\noindent
Using the master symmetry $X_1$ and properties of the Schouten bracket one
can derive  similar relations between $\pi_2$ and $\pi_3$.
  For example   $ \pi_2 { \rm grad} \,  M_1 = \pi_3 {\rm grad} \, K_1 $, where
$M_1=  K_2 + { 1 \over 2} K_1^2= { 1 \over 2} X_1 (K_1)$.
These Lenard-type relations can be
used to prove involution of integrals.   In the course of the proof one uses the fact
 that $K_3$ 
 is a  Casimir for $\pi_2$ and $K_4$ is a  Casimir for  both  $\pi_2$ and $\pi_3$.

\bigskip
 {\bf Acknowledgments.}  Part of this work was done at the University of Arizona under the supervision of
 Hermann Flaschka. I would like to
thank him for  introducing me to this area of Mathematics and for his useful  ideas and 
suggestions.

\end{document}